\global\def\draftcontrol{0}

   \def\versionno{ lambda1 from the boost invariant expansion -- draft   }
\catcode`\@=11

\expandafter\ifx\csname draftcontrol\endcsname\relax\global\def\draftcontrol{0}
\fi

{\count255=\time\divide\count255 by 60
\xdef\hourmin{\number\count255}
\multiply\count255 by-60\advance\count255 by\time
\xdef\hourmin{\hourmin:\ifnum\count255<10 0\fi\the\count255}}
\def\draftdate{\number\month/\number\day/\number\year\ \ \ \hourmin }

\newcommand\makepapertitle{\par
  \begingroup
    \renewcommand\thefootnote{\@fnsymbol\c@footnote}%
    \def\@makefnmark{\rlap{\@textsuperscript{\normalfont\@thefnmark}}}%
    \long\def\@makefntext##1{\parindent 1em\noindent
            \hb@xt@1.8em{%
                \hss\@textsuperscript{\normalfont\@thefnmark}}##1}%
     \newpage
     \global\@topnum\z@   % Prevents figures from going at top of page.
     \@makepapertitle
     \thispagestyle{empty}\@thanks
  \endgroup
  \setcounter{footnote}{0}%
  \global\let\thanks\relax
  \global\let\makepapertitle\relax
  \global\let\@makepapertitle\relax
  \global\let\@thanks\@empty
  \global\let\@author\@empty
  \global\let\@date\@empty
  \global\let\@title\@empty
  \global\let\title\relax
  \global\let\author\relax
  \global\let\date\relax
  \global\let\and\relax
  \def\version{\let\version\@version\@gobble}
}
\def\@makepapertitle{%
  \newpage
   \ifnum\draftcontrol=1 {}
   \version\versionno
   \vskip 3em%
   \else
   \hfill\hbox to 3cm {\parbox{4cm}{\@pubnum}\hss}%
   \vskip 3em%
   \fi
   \begin{center}%
   \let \footnote \thanks
     {\LARGE {\@title}}%
     \vskip 1.5em%
     {\normalsize%\large
       \lineskip .5em%
       \begin{tabular}[t]{c}%
         \@author
       \end{tabular}\par}%
     \vskip 1.5em%
     {\@bstract}%
     \end{center}%
     \vskip 1.5em
     \@date%
   \par
}

\gdef\@pubnum{}
\def\pubnum#1{%
  \gdef\@pubnum{#1}}

\gdef\@bstract{}
\def\Abstract#1{%
  \gdef\@bstract{%
   \parbox{\textwidth-0pc}{%
   \centerline{\bf Abstract}\penalty1000%
\kern.2cm%
\noindent%\abstractfont \baselineskip=12pt
\renewcommand\baselinestretch{1.0}%
{#1}}}
}

\def\ps@paper{\let\@mkboth\@gobbletwo%
     \ifnum\draftcontrol=1
    \def\@oddfoot{\hbox to \textwidth{\tiny \versionno \hfil\tiny\draftdate}%
    \hskip -\textwidth \hbox to \textwidth{\hfil\rm\thepage\hfil}}%
     \else\def\@oddfoot{\hbox to \textwidth{\hfil\rm\thepage\hfil}}
     \fi
     \let\@evenfoot\@oddfoot
}
\def\body{\clearpage
          \pagestyle{paper}
    }
\def\@version#1{\ifnum\draftcontrol=1
\typeout{}\typeout{#1}\typeout{}
\vskip3mm\centerline{\hbox{\fbox{\normalsize{\tt DRAFT -- #1 -- }
                   {\draftdate}}}}\vskip3mm
\fi}
\let\version\@version
\long\def\eqlabel#1{\ifnum\draftcontrol=1
                    \tag@false  % there are some problems with multline without this
                    \tag*{(\theequation) \hbox to -0.2cm{\hspace{0cm}\small{#1}\hss}}
                    \refstepcounter{equation}
                    \edef\@currentlabel{\theequation}
                    \ltx@label{#1}          % use old LaTeX \label instead of new definition
                    \else
                    \label{#1}
                    \fi
                    }
\let\st@bibitem\@bibitem
\let\st@lbibitem\@lbibitem
\ifnum\draftcontrol=1
  \def\@bibitem#1{%
    \st@bibitem{#1}\a@@label{#1}\ignorespaces}
  \def\@lbibitem[#1]#2{%
    \st@lbibitem[#1]{#2}\a@@label{#2}\ignorespaces}
  \def\a@@label#1{%
    \gdef\a@lab{\smash{\normalfont\small#1}}
    \ifvmode
      \if@inlabel
        \global\setbox\@labels\hbox{%
          \llap{\a@lab\let\a@lab\relax
                \kern\@totalleftmargin\kern\marginparsep}%
          \box\@labels}%
      \fi
    \fi}
\fi
\documentclass[12pt,letterpaper]{article}
\usepackage{amsmath,amssymb,array,calc,epsfig}
\ifnum\draftcontrol=1
\fi
\renewcommand\baselinestretch{1.25}
\renewcommand\section{\@startsection {section}{1}{\z@}%
                                   {-3.5ex \@plus -1ex \@minus -.2ex}%
                                   {2.3ex \@plus.2ex}%
                                   {\normalfont\large\bfseries}}
\renewcommand\subsection{\@startsection{subsection}{2}{\z@}%
                                   {-3.25ex\@plus -1ex \@minus -.2ex}%
                                   {1.5ex \@plus .2ex}%
                                   {\normalfont\normalsize\bfseries}}
\renewcommand\subsubsection{\@startsection{subsubsection}{3}{\z@}%
                                   {-3.25ex\@plus -1ex \@minus -.2ex}%
                                   {1.5ex \@plus .2ex}%
                                   {\normalfont\normalsize\it}}
\renewcommand\paragraph{\@startsection{paragraph}{4}{\z@}%
                                   {-3.25ex\@plus -1ex \@minus -.2ex}%
                                   {1.5ex \@plus .2ex}%
                                   {\normalfont\normalsize\bf}}

\numberwithin{equation}{section}

\def\ie{{\it i.e.}}

\def\revise#1       {\raisebox{-0em}{\rule{3pt}{1em}}%
                     \marginpar{\raisebox{.5em}{\vrule width3pt\
                     \vrule width0pt height 0pt depth0.5em
                     \hbox to 0cm{\hspace{0cm}{%
                     \parbox[t]{4em}{\raggedright\footnotesize{#1}}}\hss}}}}

\def\cala         {{\cal A}}

\def\calc         {{\cal C}}

\def\cali         {{\cal I}}
\def\calj         {{\cal J}}

\def\caln         {{\cal N}}
\def\calo         {{\cal O}}

\def\calq         {{\cal Q}}
\def\calr         {{\cal R}}
\def\cals         {{\cal S}}

\def\del          {\partial}

\def\sqr#1#2{{\vcenter{\vbox{\hrule height.#2pt
 \hbox{\vrule width.#2pt height#1pt \kern#1pt
 \vrule width.#2pt}\hrule height.#2pt}}}}

\def\a{\alpha}

\def\r{\rho}
\def\dd{\delta}
\def\e{\epsilon}

\def\ga{\gamma}

\def\aa1{\phi}
\def\cc1{\psi}

\def\arctanh{{\rm arctanh}}
\def\t{\tau}

\def\l{\lambda}
\def\ha{\hat{a}}
\def\hb{\hat{b}}
\def\hc{\hat{c}}
\def\hal{\hat{\alpha}}
\def\hp{\hat{\phi}}
\def\he{\hat{\eta}}

\newcommand{\eq}{\begin{equation}}
\newcommand{\eqx}{\end{equation}}
\newcommand{\eqn}{\begin{eqnarray}}
\newcommand{\eqnx}{\end{eqnarray}}

\catcode`\@=12

\begin{document}

%%%
%%%%%% text starts here
%%%%%%%%%

\title{\bf Second order hydrodynamics of a CFT plasma from boost invariant expansion}
\pubnum{%
UWO-TH-08/13
DAMTP-2008-67}

\date{August 2008}

\author{
Alex Buchel$ ^{1,2}$ and Miguel Paulos$ ^3$\\[0.4cm]
\it $ ^1$Department of Applied Mathematics\\
\it University of Western Ontario\\
\it London, Ontario N6A 5B7, Canada\\[0.2cm]
\it $ ^2$Perimeter Institute for Theoretical Physics\\
\it Waterloo, Ontario N2J 2W9, Canada\\[0.2cm]
\it $ ^3$Department of Applied Mathematics and \\
\it Theoretical Physics, Cambridge CB3 0WA, U.K.
    }

\Abstract{
We compute finite coupling correction to a nonlinear second order
hydrodynamic coefficient in the boost invariant expansion of the
$\caln=4$ supersymmetric Yang-Mills plasma. The result is universal
for a large class of strongly coupled four dimensional conformal gauge
theories.
}

\makepapertitle

\body

\version\versionno

\section{Introduction and summary}
It is common to model the boost-invariant expansion  \cite{bj} of the strongly coupled (relativistic) quark-gluon plasma (sQGP) produced in heavy ion 
collisions at RHIC in the framework of Mueller-Israel-Stewart theory (MIS) \cite{m,is}. 
One often further approximates sQGP dynamics in the relevant regime as that of a conformal theory.
For a four-dimensional viscous conformal plasma undergoing 
boost-invariant expansion, MIS theory predicts the evolution of the energy density $\e(\t)$  and the component of the viscous flow 
$\Phi(\t)$ as (see \cite{mur})   
\begin{equation}
\begin{split}
\del_\t\e=&-\frac {4\e}{3\t}+\frac\Phi\t\,,\\
\t_\Pi\del_\t\Phi=&\frac{4\eta}{3\t}-\Phi-\frac{4\t_\Pi}{3\t}\Phi\,,
\end{split}
\eqlabel{mist}
\end{equation}
where the two phenomenological parameters are: $\eta\propto T^3$, the plasma shear viscosity, and $\tau_\Pi\propto T^{-1}$, the plasma relaxation time. 

To a large extent motivated and guided by the gauge theory/string theory correspondence of Maldacena 
\cite{m9711,m2,ss}, Baier {\it et.al} \cite{rel1} and Bhattacharyya {\it et.al} \cite{rel2}  recently formulated a complete theory of 
the second order relativistic viscous hydrodynamics of conformal fluids. In this theory, the MIS linearized equations governing the boost-invariant 
expansion \eqref{mist} are modified by the inclusion of a new  term, quadratic in the component $\Phi$ of the viscous flow:
\begin{equation}
\begin{split}
\del_\t\e=&-\frac {4\e}{3\t}+\frac\Phi\t\,,\\
\t_\Pi\del_\t\Phi=&\frac{4\eta}{3\t}-\Phi-\frac{4\t_\Pi}{3\t}\Phi-\frac {\lambda_1}{2\eta^2}\Phi^2\,,
\end{split}
\eqlabel{mistn}
\end{equation}
where $\lambda_1\propto T^2$ is a new second order hydrodynamic coefficient.
An important observation of \cite{rel1,rel2} was that for the $\caln=4$ supersymmetric Yang-Mills plasma (SYM), and in fact for 
all (infinitely) strongly coupled conformal gauge theory plasmas, allowing for a dual string theory description, 
$\lambda_1\ne 0$. Moreover, for the near-equilibrium dynamics\footnote{The near-equilibrium dynamics corresponds to the late-time 
boost-invariant expansion (see \cite{bbe}).}
 (where one can reasonably apply hydrodynamics at  all)   
the nonlinear term in \eqref{mistn} is equally important to $\tau_\Pi$ terms of the MIS theory, introduced to restore causality in
first-order hydrodynamics. As reported in \cite{rel1,rel2}, and for the shear viscosity in \cite{s0},  any (infinitely) 
strongly coupled four-dimensional conformal gauge theory plasma (with a string theory dual) has
\begin{equation}
\begin{split}
\frac{\eta}{s}=\frac{1}{4\pi}\,,\qquad \tau_\Pi T=\frac{2-\ln 2}{2\pi}\,,\qquad \frac{\lambda_1T}{\eta}=\frac{1}{2\pi}\,,
\end{split}
\eqlabel{sugravalues}
\end{equation}
where $s$ is the entropy density.

The finite coupling corrections to $\eta/s$  and $\t_\Pi T$ for $\caln=4$ SYM plasma were 
computed
 in \cite{bls,bb,bmis,beta,p}
and \cite{bp}:
\begin{equation}
\begin{split}
\frac{\eta}{s}=\frac{1}{4\pi}\left(1+\frac{120}{8}\zeta(3)\lambda^{-3/2}+\cdots\right)\,,\qquad \t_\Pi T=\frac{2-\ln 2}{2\pi}
+\frac{375}{32\pi}\zeta(3)\lambda^{-3/2}+\cdots\,,
\end{split}
\eqlabel{corrold}
\end{equation}
where $\lambda$ is the gauge theory 't Hooft coupling.
It was proposed in \cite{bun}\footnote{The proof is given in \cite{bmps}.} that corrections \eqref{corrold} are universal 
for all  four dimensional conformal gauge theory plasmas (with equal $a$ and $c$ central charges) that allow for  string theory duals.  
Similarly, within the same class of conformal plasmas, the finite coupling correction to $\lambda_1$ is universal.

In this paper we report the (universal) finite coupling correction to $\lambda_1$. We find
\begin{equation}
\begin{split}
\frac{\lambda_1T}{\eta}=\frac{1}{2\pi}\left(1+\frac{215}{8}\zeta(3)\lambda^{-3/2}+\cdots\right)\,.
\end{split}
\eqlabel{corrnew}
\end{equation}
Given \eqref{corrold} and \eqref{corrnew} we have now a complete set of universal phenomenological parameters 
describing boost-invariant expansion of conformal gauge theory plasmas at finite 't Hooft coupling. 
We hope these results will prove useful in numerical hydrodynamic simulations of RHIC (and LHC) nuclear collisions.

The computations are quite technical, so we present only relevant steps and for the details refer the reader 
to previous work on the subject. We rely on the pioneering work of Janik and Peschanski \cite{j1} 
(and important further developments in \cite{j15,j2,hj}) which sets up a  to study boost invariant 
expansion of a gauge theory plasma in a dual string theory setting. We use notations and results (often 
without quoting them here due to their technical nature) of \cite{beta}. Some further technical details appear in 
Appendices \ref{background} and \ref{riem2}, and  supplemental data is available as \cite{sources}. 

We would like to comment on the issue of singularities in the Janik-Peschanski framework. 
To extract $\lambda_1$ we need to go to the third order in the late proper time expansion of the 
dual string theory background \cite{hj,rel1}. This is the first time where the singularities 
of the dual background geometry can not be completely removed by appropriately adjusting the hydrodynamic parameters \cite{hj,bbhj}.
Specifically, appropriate hydrodynamic parameters will remove all the pole singularities in the curvature invariants 
in the late time expansion, but the logarithmic singularities will persist in quadratic (and higher order) Riemann tensor invariants.  
It was understood in \cite{sh} (also independently in \cite{jp}) that a singular proper time redefinition in the bulk will remove all the singularities 
identified in \cite{hj,bbhj}. Since such proper time redefinition has only a logarithmic singularity, it can not affect 
the condition for the absence of pole singularities which determines the hydrodynamic parameters. Thus, we are justified to 
use the original framework of Janik-Peschanski to extract $\lambda_1$.  

Given the complexity of the computations, we feel that an independent check on the analysis is important. Thus, while computations in Secs.~2-4 are done
in the Janik-Peschanski framework, in Sec.~5 we reanalyze boost-invariant expansion of a conformal gauge theory plasma in singularity-free approach of Kinoshita {\it et al} \cite{jp}.
Both approaches lead to the same value of $\lambda_1$.

\section{Janik-Peschanski dual to a boost invariant plasma expansion}
The framework to study string theory duals of  boost-invariant plasmas was proposed in \cite{j1}. Here we closely follow notations 
and analysis in \cite{bbhj} and \cite{beta}. Most details\footnote{The details 
of the analysis are available from the authors upon request.} (including the description of the computational framework)
are omitted due to their technical nature and the fact that 
they have already been explained in \cite{beta}. 

The string theory background holographically dual to a Bjorken flow of the $\caln=4$ plasma takes the form \cite{j1,bbhj,beta}
\begin{equation}\eqlabel{10dimM}
 \begin{split}
  d\tilde{s}_{10}^{2}\:&=\:\tilde{g}_{MN}d\xi^{M}d\xi^{N}\:=\\
  &=\:e^{-2 \a(\t,z)}g_{\mu\nu}(x)dx^{\mu}dx^{\nu}+
      e^{6/5\a(\t,z)}\left(dS^{5}\right)^{2}\,,
 \end{split}
\end{equation}
 where $(dS^{5})^{2}$ is the line element for a $5$-dimensional sphere with unit
radius, and 
\begin{equation}\eqlabel{5dimM}
 \begin{split}
  ds^{2}\:&=\:g_{\mu\nu}dx^{\mu}dx^{\nu}\:=\:\\
          &=\frac{1}{z^{2}}
	    \left[
             -e^{2a(\tau,z)}d\tau^{2}+e^{2b(\tau,z)}\tau^{2}dy^{2}+
	      e^{2c(\tau,z)}dx_{\perp}^{2}
	    \right]+
	    \frac{dz^{2}}{z^{2}}\,,
 \end{split}
\end{equation}
where $dx_{\perp}^{2}\equiv dx_1^2+dx_2^2$.
The $5$-form $F_{5}$ takes form\footnote{We normalize the five-form flux so that the asymptotic AdS radius is one.}
\begin{equation}\eqlabel{5form}
F_{5}\:=\:\mathcal{F}_{5}+\star\mathcal{F}_{5}\,,\qquad
\mathcal{F}_{5}\:=\:-4\ \omega_{S^{5}}\,,
\end{equation}
where $\omega_{S^{5}}$ is the 5-sphere volume form.
Moreover,  the dilaton is $\phi=\phi(\t,z)$.

Equations of motion for the metric warp factors and the dilaton  are solved as a series expansion in the late proper time $\t\to \infty$, but exactly 
in the scaling variable \cite{j1}
\begin{equation}
v\equiv \frac{z}{\t^{1/3}}\,.
\eqlabel{vv}
\end{equation}   
Specifically,
\begin{equation}
\begin{split}
a(\tau,v) =& \biggl(a_{0}(v)+\ga \ha_0(v)\biggr) + \frac{1}{\tau^{2/3}}\biggl(a_{1}(v)+\ga \ha_1(v)\biggr) + \frac{1}{\tau^{4/3}} 
\biggl(a_{2}(v)+\ga \ha_2(v)\biggr)\\
&+ \frac{1}{\tau^{2}} 
\biggl(a_{3}(v)+\ga \ha_3(v)\biggr)\,,\\
b(\tau,v) =& \biggl(b_{0}(v)+\ga \hb_0(v)\biggr) + \frac{1}{\tau^{2/3}}\biggl(b_{1}(v)+\ga \hb_1(v)\biggr) + \frac{1}{\tau^{4/3}} 
\biggl(b_{2}(v)+\ga \hb_2(v)\biggr)\\
&+ \frac{1}{\tau^{2}} 
\biggl(b_{3}(v)+\ga \hb_3(v)\biggr)\,,\\
c(\tau,v) =& \biggl(c_{0}(v)+\ga \hc_0(v)\biggr) + \frac{1}{\tau^{2/3}}\biggl(c_{1}(v)+\ga \hc_1(v)\biggr) + \frac{1}{\tau^{4/3}} 
\biggl(c_{2}(v)+\ga \hc_2(v)\biggr)\\
&+ \frac{1}{\tau^{2}} 
\biggl(c_{3}(v)+\ga \hc_3(v)\biggr)\,,\\
\a(\tau,v) =& \ga \hal_0(v) + \frac{1}{\tau^{2/3}}\ \ga \hal_1(v) + \frac{1}{\tau^{4/3}} 
\ga \hal_2(v)+ \frac{1}{\tau^{2}} 
\ga \hal_3(v)\,,\\
\phi(\tau,v) =& \ga \hp_0(v) + \frac{1}{\tau^{2/3}}\ \ga \hp_1(v) + \frac{1}{\tau^{4/3}} 
\ga \hp_2(v)+ \frac{1}{\tau^{2}} 
\ga \hp_3(v)\,,
\end{split}
\eqlabel{defform}
\end{equation}
where 
\begin{equation}
\ga=\frac 18 \zeta(3)\ \left(\a'\right)^3\qquad \Leftrightarrow\qquad \frac 18 \zeta(3)\ \lambda^{-3/2} 
\eqlabel{defga}
\end{equation}
is the leading string theory $\a'$-correction to type IIB supergravity.
As argued in \cite{p},\cite{mp}, the five-form  is not corrected to this order.

Solutions for the background warp factors $\{a_i,b_i,c_i\}$, $i=0,1,2,3$  were obtained in \cite{hj}; 
the  leading string theory $\a'$-corrections, up to the second order, $\{\ha_i,\hb_i,\hc_i,\hal_i,\hp_i\}$ with $i=0,1,2$,
were discussed in \cite{beta}. In the next section, we extended the analysis of \cite{beta} to the leading string theory 
$\a'$-corrections to the supergravity background at the third order, \ie, for $\ha_3$, $\hb_3$, $\hc_3$, $\hal_3$, $\hp_3$.

\section{Equations of motion and solutions for $\{\ha_3,\hb_3,\hc_3,\hal_3,\hp_3\}$}

We obtain equations of motion (including the constraints) at the third order  for  $\ha_3$, $\hb_3$, $\hc_3$, $\hal_3$, $\hp_3$, 
extending the analysis in \cite{beta}.  
All the equations must be solved with the boundary conditions
\begin{equation}
\biggl\{\ha_3(v),\hb_3(v),\hc_3(v),\hal_3(v),\hp_3(v)\biggr\}\bigg|_{v\to 0}=0\,.
\eqlabel{bbcc}
\end{equation}

We find the following set of equations for the next-to-next-to-next-to-leading 
order in the late proper time expansion at order $\calo(\ga)$
\begin{equation}
\begin{split}
0=&\hc_3''+\frac 12 \hb_3''+\frac{5 v^4-9}{(3+v^4) v} \hc_3'+\frac {5 v^4-9}{2(3+v^4) v} \hb_3'+\cals_{(3,1)}\,,
\end{split}
\eqlabel{eom31}
\end{equation} 
\begin{equation}
\begin{split}
0=&\hc_3''+\frac 12 \ha_3''+\frac{5 v^8+27}{v  (v^8-9)} \hc_3'+ \frac{9+5 v^4}{2v (v^4-3)} \ha_3'+\cals_{(3,2)}\,,
\end{split}
\eqlabel{eom32}
\end{equation} 
\begin{equation}
\begin{split}
0=&\hc_3''+\ha_3''+\hb_3''+\frac{9+5 v^4}{v (v^4-3)} \ha_3'+\frac{5 v^8+27}{v  (v^8-9)} \hb_3'+\frac{5 v^8+27}{v (v^8-9)} \hc_3'+\cals_{(3,3)}\,,
\end{split}
\eqlabel{eom33}
\end{equation} 
\begin{equation}
\begin{split}
0=&\hc_3''+\frac 12 \hb_3''- \frac{3(v^4-3)}{2(3+v^4) v} \ha_3'+\frac{3(v^8-5 v^4-6)}{v  (v^8-9)} \hb_3'+\frac{3(3 v^8-10 v^4-21)}{v (v^8-9)} \hc_3'
-\frac{72 v^2}{v^8-9} \hb_3\\
&-\frac{144 v^2}{v^8-9} \hc_3+\cals_{(3,4)}\,,
\end{split}
\eqlabel{eom34}
\end{equation} 
\begin{equation}
\begin{split}
0=&\hc_3'+\frac 12 \hb_3'+\frac{(v^4-3)^2}{2(v^4-2 v^2+3) (v^4+2 v^2+3)} \ha_3'+\cals_{(3,5)}\,,
\end{split}
\eqlabel{eom35}
\end{equation}
\begin{equation}
\begin{split}
0=&\hal_3''+\frac{5 v^8+27}{v  (v^8-9)} \hal_3'-\frac{32}{v^2} \hal_3+\cals_{(3,6)}\,,
\end{split}
\eqlabel{eom36}
\end{equation} 
\begin{equation}
\begin{split}
0=&\hp_3''+\frac{5 v^8+27}{v (v^8-9)} \hp_3'+\cals_{(3,7)}\,,
\end{split}
\eqlabel{eom37}
\end{equation} 
where the source terms $\{\cals_{(3,1)}\cdots,\cals_{(3,7)}\}$ are given in \cite{sources}.
While the system \eqref{eom31}-\eqref{eom37} is overdetermined, we explicitly verified that 
it is consistent. 

Solving \eqref{eom31}-\eqref{eom37} is quite complicated. Fortunately, we do not need a complete solution. 
Our ultimate goal is to determine $C$ from the nonsingularity of the ten dimensional metric curvature invariants
to order $\calo(\ga)$ and to order $\calo(\t^{-2})$ in the late proper time expansion. Thus we evaluate 
metric invariants first, find  what field combinations affect the singularity as $v\to 3^{1/4}_-$, and then solve just for 
those combinations of fields.
  
We assume  that 
\begin{equation}
\he=\he_0+\ga \he_1+\calo(\ga^2)\,,\qquad C=C_0+\ga C_1+\calo(\ga^2)\,,
\eqlabel{etaex}
\end{equation}
and evaluate background curvature invariants to order $\calo(\ga)$ near 
\begin{equation}
x\equiv 3^{1/4}-v\,.
\eqlabel{ydef}
\end{equation} 
We use explicit solutions at lower orders, as well as equations of motion for the second and the third order  
to eliminate the derivatives (if possible) of $$\{a_3,b_3,c_3;\,\,\,\ha_2,\hb_2,\hc_2,\hal_2;\,\,\,\ha_3,\hb_3,\hc_3,\hal_3\}$$ 
from the curvature invariants (see \cite{beta} for details).

\subsection{$\calr$ at order $\calo(\t^{-2})$}
For the  Ricci scalar we find 
\begin{equation}
\begin{split}
\calr=&\cdots+\frac{\ga}{\t^2 (v^4+2 v^2+3) (v^4-2 v^2+3) (3+v^4)^3} \biggl\{ -\frac{207360 v^{13} (v^4-3)^3}{(3+v^4)^4} a_3'\\
&-\frac{497664\he_0 v^{13}}{(v^4+2 v^2+3) (v^4-2 v^2+3) (v^8-9) (3+v^4)^4} \biggl(2 v^{28}-56 v^{24}+89 v^{20}-765 v^{16}\\
&+900 v^{12}-9882 v^8+7857 v^4-6561\biggr)  
a_2'
-\frac{13824 v^{14} \he_0}{(v^8-9)^4 (3+v^4)} \biggl(25 v^{28}+324 \he_0^2 v^{26}\\
&+189 v^{24}+6696 v^{22} \he_0^2+1269 v^{20}+69660 \he_0^2 v^{18}+7641 v^{16}
-58320 v^{14} \he_0^2+22923 v^{12}\\
&+626940 \he_0^2 v^{10}+34263 v^8+542376 \he_0^2 v^6+45927 v^4+236196 \he_0^2 v^2\\
&+54675\biggr)\ln\frac{3-v^4}{3+v^4}
+\frac{1990656 \he_0 v^{16}(b_2+2 c_2)}{(v^8-9)^4 (3+v^4)} \biggl(3 v^{24}+62 v^{20}+645 v^{16}-540 v^{12}\\
&+5805 v^8+5022 v^4+2187\biggr)
-\frac{1492992v^{16} \he_0^3}{(v^4-2 v^2+3) (v^4+2 v^2+3) (3+v^4)^3 (v^8-9)^5}\\
&\times \biggl(37 v^{48}+4434912 v^{24}+14348907+1273320 v^{20}
+7452 v^{36}+315801 v^{32}\\
&+46008 v^{28}+17684 v^{40}-36137988 v^4-300 v^{44}+12538071 v^{16}-28072332 v^{12}\\
&+78180876 v^8
\biggr)
+\frac{27648 \he_0 v^{18}}{(v^4-2 v^2+3) (v^4+2 v^2+3) (v^8-9)^5} \biggl(53 v^{36}-533 v^{32}\\
&+6978 v^{28}+4518 v^{24}+123228 v^{20}+112428 v^{16}+1289358 v^{12}+1255338 v^8\\
&+693279 v^4
+4153113\biggr)\biggr\}\,,
\end{split}
\eqlabel{rsc2}
\end{equation}
where $\cdots$ denote lower orders in the later time expansion.
We recall explicit expressions for $\{a_2,b_2,c_2\}$ as well as present the decoupled equation for $a_3$ in Appendix \ref{background}. 
Using \eqref{order2}, \eqref{expa3} we find
\begin{equation}
\begin{split}
\calr=&\cdots+\frac{\ga}{\t^2}\biggl\{-3 \he_0 3^{1/4} \left(\sqrt3-18 \he_0^2\right)\ \frac{1}{x^5}+\frac 34 \he_0 
\left(13 \sqrt3-4 \ln2 \sqrt3-30 \he_0^2+2 C_0\right)\ \frac{1}{x^4}\\
&-\frac 14 \he_0 3^{1/4} \left(426 \he_0^2 \sqrt3+4 C_0 \sqrt3-3-24 \ln2\right)\ \frac{1}{x^3}\\
&-\frac 14 \he_0 \left(75-78 \ln2+876 \he_0^2 \sqrt3+13 C_0 \sqrt3\right)\ \frac{1}{x^2}\\
&+\frac{15}{4} \he_0 3^{1/4} \left(\sqrt3-2 \ln2 \sqrt3+84 \he_0^2+C_0\right)\ \frac 1x-64\sqrt 3\he_0\ \ln x+{\rm finite}\biggr\}\,.
\end{split}
\eqlabel{rsc3}
\end{equation}
From \eqref{rsc3}  we  find that the  Ricci scalar of the string theory geometry does not have pole singularities as $x\to 0_+$ when 
\begin{equation}
\he_0=\frac{1}{2^{1/2}3^{3/4}}\,, \qquad C_0=2\sqrt 3 \ln 2-\frac{17}{\sqrt 3}\,,
\eqlabel{eta0}
\end{equation}
which are precisely the conditions found from the nonsingularity of the Riemann tensor squared \cite{j2,hj}, 
as well as higher curvature invariants \cite{bbhj},
in the supergravity approximation to the string theory dual of the $\caln=4$ SYM Bjorken flow. The difference here (compare to \cite{j2,hj,bbhj}) 
is that $\he_0$ and $C_0$ are already fixed by requiring the nonsingularity of the Ricci scalar.

While the pole singularities in the ten dimensional Ricci scalar are removed, given \eqref{eta0},  the logarithmic singularity 
still persists. A similar observation was made at the supergravity level for the Riemann tensor invariants in \cite{hj,bbhj}.
We expect that the remaining logarithmic singularity is removed by an appropriate change of variable \cite{sh,jp}.
 
\subsection{$\calr_{\mu\nu\r\l}\calr^{\mu\nu\r\l}$ at order $\calo(\t^{-2})$}
A bit more work is necessary to determine the nonsingularity condition of the Riemann tensor squared at order $\calo(\ga)$.
Generalizing the notation of \cite{bbhj} 
\begin{equation}
\begin{split}
\cali^{[2]}&\equiv \calr_{\mu\nu\r\l}\calr^{\mu\nu\r\l}\\
&=\biggl(\cali^{[2]SUGRA}_0(v)+\ga\ \cali^{[2]W}_0(v)\biggr)
+\frac{1}{\t^{2/3}} \biggl(\cali^{[2]SUGRA}_1(v)+\ga\ \cali^{[2]W}_1(v)\biggr)
\\
&+\frac{1}{\t^{4/3}} \biggl(\cali^{[2]SUGRA}_2(v)+\ga\ \cali^{[2]W}_2(v)\biggr)+\frac{1}{\t^{2}} \biggl(\cali^{[2]SUGRA}_3(v)
+\ga\ \cali^{[2]W}_3(v)\biggr)\\
&+\calo(\t^{-8/3})+\calo(\ga^2)\,.
\end{split}
\eqlabel{i3def}
\end{equation}

Explicitly we find: 
\begin{equation}
\begin{split}
&\cali^{[2]SUGRA}_3=\frac{1}{(v^4+2 v^2+3) (v^4-2 v^2+3)} \biggl\{\frac{1152 v^5 (v^4-3)^3}{(3+v^4)^3} a_3'
\\
&-\frac{13824v^5 \he}{(3+v^4)^3 (v^8-9) (v^4+2 v^2+3) (v^4-2 v^2+3)} \biggl(8 v^{24}+17 v^{20}+133 v^{16}+210 v^{12}\\
&+1674 v^8-243 v^4+729\biggr) a_2'
-\frac{55296 \he v^8}{(v^8-9)^4} \biggl(v^{24}+18 v^{20}+183 v^{16}+60 v^{12}+1647 v^8\\
&+1458 v^4+729\biggr) (b_2+2 c_2)+\frac{384 v^6 \he}{(v^8-9)^4} \biggl(5 v^{28}+108 \he^2 v^{26}+57 v^{24}+1944 v^{22} \he^2\\
&+465 v^{20}+19764 \he^2 v^{18}
+2565 v^{16}+6480 v^{14} \he^2+7695 v^{12}+177876 \he^2 v^{10}+12555 v^8\\
&+157464 \he^2 v^6+13851 v^4+78732 \he^2 v^2+10935\biggr) 
\ln\frac{3-v^4}{3+v^4}
\\
&+\frac{82944 v^{12}\he^3}{(v^8-9)^5 (v^4-2 v^2+3) (v^4+2 v^2+3) (3+v^4)^2} \biggl(434268 v^{20}+794286 v^{16}+514674 v^{12}\\
&+2322594 v^4
-334611 v^8-531441+114642 v^{24}+37740 v^{28}+49 v^{40}+2 v^{44}+1314 v^{36}\\
&+7251 v^{32}\biggr) -\frac{768\he v^{10}}{(v^4-3) (v^4-2 v^2+3) (v^4+2 v^2+3) (v^8-9)^4} \biggl(v^{36}-25 v^{32}+1026 v^{28}
\\
&+6054 v^{24}+33300 v^{20}+88452 v^{16}+338094 v^{12}+392202 v^8+317115 v^4+452709\biggr)\biggr\}\,,
\end{split}
\eqlabel{i3sugra}
\end{equation}
\begin{equation}
\begin{split}
\cali^{[2]W}_3=&\frac{1152 v^5 (v^4-3)^3}{(v^4+2 v^2+3) (v^4-2 v^2+3) (3+v^4)^3}\ha_3'-\frac{192}{(3+v^4)^4} \biggl(
5 v^{16}+60 v^{12}+54 v^8\\
&+540 v^4+405\biggr) \hal_3
-\frac{13824v^5 \he_0}{(v^4+2 v^2+3)^2 (v^4-2 v^2+3)^2 (v^4+3)^3 (v^8-9)} \biggl(8 v^{24}\\
&+17 v^{20}+133 v^{16}+210 v^{12}+1674 v^8-243 v^4+729\biggr) \ha_2'
\\
&-\frac{55296 \he_0 v^8}{(v^8-9)^4 (v^4+2 v^2+3) (v^4-2 v^2+3)} \biggl(v^{24}+18 v^{20}+183 v^{16}+60 v^{12}\\
&+1647 v^8+1458 v^4+729\biggr) 
(\hb_2+2 \hc_2)+\frac{165888 \he_0 v^8 (v^4-3)}{(3+v^4)^5} \hal_2\\
&-\frac{192 v^5 \calq_{1}}{(3+v^4)^9 (v^4+2 v^2+3)^2 (v^4-2 v^2+3)^2}a_3'
\\
&+\frac{768 v^5 \calq_{2}}{(3+v^4)^7 (v^8-9)^3 (v^4+2 v^2+3)^3 (v^4-2 v^2+3)^3}a_2' 
\\
&+\frac{3072 \calq_3 v^8}{(3+v^4)^4 (v^8-9)^6 (v^4+2 v^2+3)^2 (v^4-2 v^2+3)^2} (b_2+2 c_2)
\\
&+\frac{64 \calq_4 v^6}{3 (3+v^4)^4 (v^8-9)^6 (v^4+2 v^2+3)^2 (v^4-2 v^2+3)^2}\ln\frac{3-v^4}{3+v^4}
\\
&+\frac{128 \calq_5 v^{10} \he_0\delta_1}{(3+v^4)^2 (v^8-9)^6 (v^4+2 v^2+3)^3 (v^4-2 v^2+3)^3}
\\
&-\frac{128 \calq_6 v^{10}\delta_2}{3 (v^8-9)^5 (v^4-2 v^2+3)^2 (v^4+2 v^2+3)^2 (3+v^4)^2}
\\
&-\frac{1492992 v^{12} \calq_7\he_0^3}{(v^4-2 v^2+3)^3 (v^4+2 v^2+3)^3 (3+v^4)^5 (v^8-9)^8}\\
& +\frac{4608 \he_0 v^{10}\calq_8}{
(v^4-2 v^2+3)^3 (v^4+2 v^2+3)^3 (3+v^4)^2 (v^8-9)^8}\,,
\end{split}
\eqlabel{i3w}
\end{equation}
where $\{\calq_1,\cdots \calq_8\}$ are given in Appendix \ref{riem2}.
Appendix \ref{riem2} also contains explicit expressions for $\ha_2'$, ${f}_2\equiv \hc_2+\frac 12\hb_2$
\cite{beta}, and the equation of motion for $\ha_3$. As in the case of $\hal_2$, although we can not 
explicitly solve for $\hal_3$, we can argue that $\hal_3(v)$
can be chosen to be  finite (along with its first derivative) as $v\to 3^{1/4}_-$, while having a vanishing nonnormalizable mode as 
$v\to 0_+$.

We now have all the necessary ingredients to determine $\he_1, C_1$ from the nonsingularity of  
$\left(\cali^{[2]SUGRA}_3(v)+\ga\cali^{[2]W}_3(v)\right)$: using results of Appendices \ref{background} and \ref{riem2},
as well as \eqref{eta0}, we find
\begin{equation}
\begin{split}
&\cali^{[2]SUGRA}_3+\ga\cali^{[2]W}_3\bigg|_{x\equiv 3^{1/4}-v\to 0_+}=\frac{\sqrt 2}{18}  \ga 
\biggl(7182+4 \sqrt 2 3^{3/4} \delta_2-15 \delta_1-72 \sqrt 2 3^{3/4} \he_{1}\biggr)\frac{1}{x^5}\\
&-\frac{1}{24} \ga \biggl(4 \sqrt 2 3^{3/4} \delta_1 \ln2-1272 \sqrt 2 3^{3/4} \ln2+30090 \sqrt 2 3^{3/4}
+24 \sqrt 3 \delta_2-21 \sqrt 2 3^{3/4} \delta_1\\
&+8 3^{1/4}\sqrt 2  C_1
-176 \sqrt 3 \he_{1}+32 \sqrt 2 3^{1/4} \delta_3\biggr)\frac{1}{x^4}+\frac{ 3^{1/4}}{72} \ga \biggl(24 \delta_2+1104 \he_{1}
+16 \sqrt 2 3^{3/4} C_1\\
&-51 \sqrt 2 3^{1/4} \delta_1-7632 \sqrt 2 3^{1/4} \ln2+144630 \sqrt 2 3^{1/4}
+64 \sqrt 2 3^{3/4} \delta_3+24 \sqrt 2 3^{1/4} \delta_1 \ln2\biggr)\frac{1}{x^3}\\
&+\frac{1}{36} \ga \biggl(-4 \delta_2+264 \he_{1}+2 \sqrt 2 3^{3/4} C_1+22 \sqrt 2 3^{1/4} \delta_1
-954 \sqrt 2 3^{1/4} \ln2+22896 \sqrt 2 3^{1/4}\\
&+8 \sqrt 2 3^{3/4} \delta_3+3 \sqrt 2 3^{1/4} \delta_1 \ln2\biggr)
\frac{1}{x^2}
-\frac{3^{1/4}}{36}  \ga \biggl(-9 \sqrt 2 3^{3/4} \delta_1 \ln2-4410 \sqrt 2 3^{3/4} \ln2\\
&+30384 \sqrt 2 3^{3/4}+44 \sqrt 3 \delta_2+24 \sqrt 2 3^{3/4} \delta_1+6 \sqrt 2 3^{1/4} C_1+264 \sqrt 3 \he_{1}+24 \sqrt 2 3^{1/4} \delta_3\biggr)\frac1x\\
&+\biggl(8 \sqrt 2 3^{3/4}-\frac29 \left(-3 \sqrt 2 3^{3/4} \delta_1-49975 \sqrt 2 3^{3/4}-216 \sqrt 3 \he_{1}+12 \sqrt 3 \delta_2\right) 
\ga\biggr) \ln x
+{\rm finite}\,.
\end{split}
\eqlabel{sing}
\end{equation}
The residues of all pole singularities in \eqref{sing} vanish provided\footnote{The value of $\he_1$ agrees with the 
one determined from the nonsingularity of the second order late-time curvature invariants in \cite{beta}.}
\begin{equation}
\begin{split}
\he_1=&\frac{3^{1/4}\sqrt{2}}{432} \left(7182-15\ \delta_1+2^{5/2}\ 3^{3/4}\ \delta_2\right)\,,\\
C_1=&-\frac{\sqrt2 3^{3/4}}{216} \biggl(122238 \sqrt2 3^{3/4}+18 \sqrt2 3^{3/4} \delta_1 \ln2-67 \sqrt2 3^{3/4} \delta_1+64 \sqrt3 \delta_2
\\
&-5724 \sqrt2 3^{3/4} \ln2+144 \sqrt2 3^{1/4} \delta_3\biggr) \,,\\
\delta_1=&-288\,,\qquad \delta_2=-144 \sqrt2 3^{1/4}\,,
\end{split}
\eqlabel{eta01}
\end{equation}
where we kept explicit dependence on $\delta_1\,, \delta_2$ in $\he_1$ and $C_1$ (as determined by the vanishing of the residues of the poles in 
\eqref{sing}
up to order three inclusive); the vanishing of the residues of the second order and the first order 
poles in \eqref{sing} determines $\delta_1$ and $\delta_2$.

As for the Ricci scalar \eqref{rsc3}, the logarithmic singularity in the Riemann tensor squared at the third order in the 
late-time expansion, \eqref{sing}, remains. This remaining singularity has both the supergravity piece (in agreement with 
\cite{hj}) and the new $\calo(\ga)$ contribution.

Notice that while $\he_0, C_0$ \eqref{eta0} are determined unambiguously from the nonsingularity condition of the 
background geometry, the absence of singularities is not a powerful enough constraint to fix $C_1$\footnote{Although, unlike the 
analysis up to the second order \cite{beta}, $\he_1$ is determined unambiguously here.}.  
This fact will not preclude us from computing a definite value of non-linear second-order hydrodynamic coefficient $\lambda_1$.

\subsection{$\calr_{\mu\nu}\calr^{\mu\nu}$ at order $\calo(\t^{-2})$}
Analysis of the square of the Ricci tensor can be performed in the same way as for the Riemann tensor squared.
We find
\begin{equation}
\calr_{\mu\nu}\calr^{\mu\nu}=\cdots+\frac{\ga}{\t^2}\biggl\{
\frac{4096}{3^{1/4}}\sqrt2 \ \ln x+ {\rm finite}-{1920\ \hal_3(x)}\biggr\}\,,\qquad x\equiv 3^{1/4}-v\to 0_+\,,
\eqlabel{ricci}
\end{equation}
where we explicitly indicated the dependence on order three fields; as before, $\cdots$ indicate lower orders in the late time expansion
studied in \cite{beta}.
We pointed out above that $\hal_3(v)$ can be chosen to be finite as $v\to 3^{1/4}_-$; this would guarantee the 
absence of pole singularities in  $\calr_{\mu\nu}\calr^{\mu\nu}$ to orders $\calo(\t^{-2})$ and $\calo(\ga)$.

\subsubsection{Higher order curvature invariants}
As in \cite{bbhj} we denote
\begin{equation}
\calr^{[2^n]}\ _{\mu\nu\r\l}\equiv \calr^{[2^{n-1}]}\ _{\mu_1\nu_1\mu\nu}\cdot \calr^{[2^{n-1}]}\ ^{\mu_1\nu_1}\ _{\r\l}\,,
\eqlabel{riemann}
\end{equation}
where
\begin{equation}
\calr^{[0]}\ _{\mu\nu\r\l}\equiv \calr_{\mu\nu\r\l}\,.
\eqlabel{n0}
\end{equation}
We further define higher curvature invariants $\cali^{[2^n]}$, generalizing \eqref{i3def}:
\begin{equation}
\begin{split}
\cali^{[2^n]}&\equiv \calr^{[2^{n-1}]}\ _{\mu\nu\r\l}\calr^{[2^{n-1}]}\ ^{\mu\nu\r\l}\\
&=\biggl(\cali^{[2^n]SUGRA}_0(v)+\ga\ \cali^{[2^n]W}_0(v)\biggr)+\frac{1}{\t^{2/3}} \biggl(\cali^{[2^n]SUGRA}_1(v)+
\ga\ \cali^{[2^n]W}_1(v)\biggr)\\
&+\frac{1}{\t^{4/3}} \biggl(\cali^{[2^n]SUGRA}_2(v)+\ga\ \cali^{[2^n]W}_2(v)\biggr)
+\frac{1}{\t^{2}} \biggl(\cali^{[2^n]SUGRA}_3(v)+\ga\ \cali^{[2^n]W}_3(v)\biggr)\\
&+\calo(\t^{-8/3})+\calo(\ga^2)\,.
\end{split}
\eqlabel{indef}
\end{equation}

Given the complexity of the analysis, we checked at order $\calo(\t^{-2})$ only the nonsingularity of $\cali^{[4]}$.
Using the results of the Appendices \ref{background} and \ref{riem2}, as well as \eqref{eta0} and \eqref{eta01}, we find
\begin{equation}
\begin{split}
&\cali^{[2]SUGRA}_3+\ga\cali^{[2]W}_3\bigg|_{x\equiv 3^{1/4}-v\to 0_+}=\biggl(
\frac{640}{3^{1/4}}\sqrt 2+\ga\frac{5636512}{3^{5/4}}\sqrt2\biggr)\ln x +{\rm finite}\,.
\end{split}
\eqlabel{i4w}
\end{equation}
The supergravity part of the logarithmic singularity in \eqref{i4w} agrees with the corresponding computation in \cite{bbhj}.

\section{$\lambda_1$ for the Bjorken flow of $\caln=4$ SYM plasma}
In the previous section we analytically evaluated $\a'$-corrected supergravity background 
dual to the Bjorken flow of $\caln=4$ SYM plasma at finite coupling to order $\calo(\t^{-2})$ 
in the late proper time expansion.  We can now extract the boundary energy density $\e(\t)$ from the one-point correlation function 
of the boundary stress energy tensor using the 
$\a'$-corrected holographic renormalization developed in \cite{bal}. We confirmed that the  final expression for the energy density can 
be evaluated as in the supergravity approximation \cite{j1,j2,bbhj}:
\begin{equation}
\e(\t)=-\frac{N^2}{2\pi^2}\ \lim_{v\to 0}\frac{2 a(v,\t)}{v^4\t^{4/3}}\,.
\eqlabel{etau}
\end{equation} 
Using the results of the lower orders in the late proper time expansion \cite{beta}, 
the details presented in Appendices \ref{background} and \ref{riem2}, \eqref{defform}, \eqref{eta0} and \eqref{eta01} 
we find\footnote{Despite the fact that at order three $\delta_1$ and $\delta_2$ are fixed (see \eqref{eta01}), we keep 
them arbitrary to compare with \cite{beta}. As in \cite{beta}, the dependence on $\delta_i$ disappears in 
physical quantities.}
\begin{equation}
\begin{split}
&\e(\t)=\frac{N^2(6+576\ \ga +\ga\ \dd_1)}{12\pi^2}\ \frac{1}{\t^{4/3}} -\frac{N^2\ 2^{1/2}\ 3^{1/4}\ (1566 \ga+8+\gamma \delta_1)}{48\pi^2}\
\frac{1}{\t^2}\\
&+\frac{N^2 3^{1/2}}{864\pi^2}\biggl(
12+24\ln2+\ga\left(2\delta_1\ln 2+\delta_1+7086+4212 \ln 2\right)\biggr)\ \frac{1}{\t^{8/3}}
+\calo(\t^{-10/3})\,.   
\end{split}
\eqlabel{etres}
\end{equation}

The string theory result \eqref{etres} should now be interpreted within second order relativistic conformal 
hydrodynamics \cite{rel1,rel2}.
For the Bjorken flow of the $\caln=4$ SYM plasma we expect \cite{rel1}
\begin{equation}
\begin{split}
&\frac{\e^{gauge}(\t)}{\calc}=\t^{-4/3}-2\eta_0\ \t^{-2}+\left[\frac32\eta_0^2-\frac23\left(
\eta_0\tau^\Pi-\lambda_1^0 
\right)
\right]\t^{-8/3}+\calo\left(\t^{-10/3}\right)\,,\\
&\eta=\calc \eta_0\left(\frac \e\calc\right)^{3/4}\,,\qquad \tau_\Pi=\tau_\Pi^0\left(\frac \e\calc\right)^{-1/4}\,,
\qquad \lambda_1=\calc\lambda_1^0\left(\frac \e\calc\right)^{1/2}\,,
\end{split}
\eqlabel{egauge}
\end{equation}
where $\calc$ is an arbitrary scale, related to the initial energy density of the expanding plasma.

To match the string theory result \eqref{etres} with \eqref{egauge} we need to recall the equation of state 
for the $\caln=4$ SYM plasma \cite{gkt}
\begin{equation}
\e(T)=\frac 38 \pi^2N^2 T^4\ \left(1+15\ga\right)\,,
\eqlabel{n4energy}
\end{equation}
and the $\caln=4$ SYM relaxation time $\t_\Pi$ \cite{bp}
\begin{equation}
\tau_\Pi T=\frac{2-\ln 2}{2\pi}+\frac{375}{4\pi}\ga\,.
\eqlabel{n4tau}
\end{equation}

Ultimately, we find:
\begin{equation}
\frac{\eta}{s}=\frac{1}{4\pi}\left(1+120\ga\right)\,,\qquad \frac{\lambda_1 T}{\eta}=\frac {1}{2\pi}\left(1
+215\ga\right)\,.
\eqlabel{lambda1}
\end{equation}
Notice that the ratio of shear viscosity to the entropy density agrees with the results
reported in \cite{bls,bb,bmis,beta}, and the supergravity part of $\lambda_1$ agrees with 
computations in \cite{rel1,rel2}. 

\section{Computation in the framework of Kinoshita \it{et al}}

In this section we compute the finite coupling correction to $\lambda_1$ by using the framework of Kinoshita and collaborators \cite{jp} for 
finding the holographic dual of an expanding plasma (see also \cite{sh}). We will work with the five-dimensional action
\begin{eqnarray}
S &=& \frac 1{16\pi G} \int d^5 x \sqrt{-g} \left(R+12+\gamma ~W\right)\,,\\
W &\equiv& -\frac 14 C_{a b c d}C^{a b}_{\ \ e f}C_{g h}^{\ \ c e}C^{g h d f}+C_{a b c d}C^{a \ c}_{\ e \ f}C_{g \ h}^{\ b \ e}C^{g d h f}\,, \label{5DAction}
\end{eqnarray}
where $C_{a b c d}$ is the five-dimensional Weyl tensor. The use of this action was justified in \cite{bmps}, where it was also shown that it leads to universal finite coupling corrections to hydrodynamic coefficients. The holographic dual to the Bjorken flow of the CFT plasma is taken to be \cite{jp} of the form
\begin{eqnarray}
ds^2&=& g_{\mu\nu}dx^{\mu}dx^{\nu}\nonumber \\
&=& -r^2 a d \tau^2+ 2d \tau dr+r^2 \tau^2 e^{2b-2c}\left(1+\frac 1{r t}\right)^2 dy^2+r^2 e^cdx_\bot^2\label{Metric}\,.
\end{eqnarray}
Plugging this form for the metric into the Einstein equations of motion, one can find the functions $a(t,r), b(t,r), c(t,r)$ order by order in a late-time expansion as before:
\begin{eqnarray}
a(t,r)&=& a_0(v)+u a_1(v)+u^2 a_2(v)+...\,, \nonumber \\
b(t,r)&=& b_0(v)+u b_1(v)+u^2 b_2(v)+...\,, \nonumber \\
c(t,r)&=& c_0(v)+u c_1(v)+u^2 c_2(v)+...\,, \nonumber
\end{eqnarray}
where $v\equiv r t^{1/3},u\equiv t^{-2/3}$. In \cite{jp} these functions were explicitly computed:

\begin{equation}
a_0(v)= 1-\frac{w^4}{v^4}\,,\qquad b_0(v)=c_0(v)=0\,;   
\end{equation}
\begin{equation}
\begin{split}
a_1(v)=& -\frac 23\frac{(\xi_1+1) v^4-3 M_1 w^4 v+w^4 \xi_1}{v^5}\,,\qquad 
b_1(v)= -\frac{\xi_1+1}{v}\,,\\
c_1(v)=& -\frac{2 \xi_1}{3 v}-\frac{1}{2} M_1 \ln \left(1-\frac{w^4}{v^4}\right)+\frac{\arctan \left(\frac{v}{w}\right)+\frac{1}{2}
   \ln \left(\frac{v-w}{v+w}\right)-\frac{\pi }{2}}{3 w}\,; \\
\end{split}
\end{equation}
\begin{equation}
\begin{split}\nonumber
&a_2(v)= -\frac{2 \left(v^4+w^4\right) \xi_2}{3 v^5}-\frac{4 \left(v^3-3 w^4 M_1\right) \xi_1}{9
   v^5}+\frac{\xi_1^2 \left(v^4-3 w^4\right)}{9 v^6}  \\
   +& \frac{\left(v^4+w^4\right) \left(9 w^2 M_1^2+1\right) \arctan
   \left(\frac{v}{w}\right)}{6 v^5 w}-\frac{\left(v^4-2 w^3 v+w^4\right) \left(9 w^2 M_1^2-1\right) \ln (v-w)}{12 v^5 w}  \\
   +& \frac{\left(v^4+2 w^3 v+w^4\right) \left(9 w^2 M_1^2-1\right) \ln (v+w)}{12 v^5 w}+\frac{\left(9 M_1^2 w^4+w^2\right)
   \ln \left(v^2+w^2\right)}{6 v^4}  \\
   -& \frac{3 (3 (12 \ln (v)+5) v M_1+4) M_1 w^4+4 \left(3 v M_2  w^4+v^3\right)}{18 v^5} \,,   \\
\end{split}
\end{equation}
\begin{equation}
\begin{split}\nonumber
b_2(v)=& -\frac{\xi_1^2}{6 v^2}+\frac{1}{4}
   M_1 \left(-24 M_1 \ln
   (v)-\frac{4}{v}+\frac{\pi }{w}\right)+\frac{\left(9 w^2 \text{$\eta_0
   $}^2-2 v M_1+1\right) \arctan
  \left(\frac{v}{w}\right)}{4 v w}  \\
  & +\frac{(3 w M_1-1)
   (2 v-3 w+3 w (4 v-3 w) M_1) \ln (v-w)}{24 v
   w^2}  \\
  &+\frac{(3 w M_1+1) (-2 v-3 w+3 w (4 v+3 w)
   M_1) \ln (v+w)}{24 v w^2}  \\
   &+\frac{1}{12} \left(18
   M_1^2+\frac{1}{w^2}\right) \ln
   \left(v^2+w^2\right)-\frac{\xi_2}{v}+\frac{1}{2 v^2}\,,
\end{split}
\end{equation}
\begin{equation}
\begin{split}
  &c_2'(v)= \frac{\left(12 w M_1 v^5-6 w v^4+\pi 
   \left(v^4-w^4\right) v+2 w^5\right) M_1 w^3}{3
   \left(v^5-v w^4\right)^2}+\frac{2 \xi_1^2}{9 v^3}  \\
   & +\frac{4
   v^2 \ln (v) M_1}{3 \left(v^4-w^4\right)}+\frac{\left(6
   \left(w^4-5 v^4\right) M_1 w^4+4 v^3
   \left(v^4+w^4\right)\right) \xi_1}{9 \left(v^5-v
   w^4\right)^2}+\frac{2 \xi_2}{3 v^2}  \\
   &+\frac{\left(v^4+3 w^4-3
   w^2 M_1 \left(4 v w^2+9 \left(v^4-w^4\right) M_1
   \right)\right) \arctan\left(\frac{v}{w}\right)}{18 v^2
   \left(v^4-w^4\right) w}  \\
   & -\frac{\left((3 w M_1-1)
   \left((v+w) \left(v^2-2 w v+3 w^2\right)-9 (v-w) w
   \left(v^2+w^2\right) M_1\right)\right) \ln (v-w)}{36
   v^2 (v-w) \left(v^2+w^2\right) w}   \\
   &    -\frac{\left((3 w M_1+1)
   \left((v-w) \left(v^2+2 w v+3 w^2\right)+9 w (v+w)
   \left(v^2+w^2\right) M_1\right)\right) \ln (v+w)}{36
   v^2 (v+w) \left(v^2+w^2\right) w}  \\
   & -\frac{\left(3 M_1
   v^3+w^2\right) \ln \left(v^2+w^2\right)}{9 v^5-9 v w^4}-\frac{\pi 
   v^3-3 w \left(4 M_2  w^4+v^2\right)}{9 \left(v^5-v w^4\right)
   w}\,.
\end{split}   
\end{equation}
Both $\xi_1$ and $\xi_2$ are gauge degrees of freedom which can be set to a convenient value. In what follows we take $\xi_1=-1$ and leave $\xi_2$ arbitrary as a cross-check on our calculations.

As explained in \cite{jp}, the energy momentum tensor of the plasma can be read off from the function $a(t,r)$, by expanding the function $a(t,r)$ in the large $r$ limit. More concretely,
\begin{eqnarray}
a(t,r)&=& 1+\frac{a^{(1)}(\tau)}{r}+...+\frac{a^{(4)}(\tau)}{r^4}+...\,,\nonumber \\
T_{\tau \tau}&\equiv &\epsilon(\tau)=-\frac 32 \frac{N^2}{4\pi^2} a^{(4)}(\tau)\,.
\end{eqnarray}
Performing this expansion in the above solution to $a(t,r)$ leads to
\begin{equation}
\epsilon(\tau)=\frac 32 \frac{N^2}{4\pi^2}\omega^4\left(\tau^{-4/3}-2 M_1 \tau^2+\left( \frac{9 M_1^2+4 M_2}6 \right)\tau^{-8/3}\right)\,.
\end{equation}
This is to be equated with the hydrodynamic expansion for $\epsilon(\tau)$,
\begin{equation}
\frac{\epsilon(\tau)}{\epsilon_0}=\tau^{-4/3}-2\eta_0 \tau^{-2}+\frac{9\eta_0^2+4(\lambda_1^0-\eta_0 \tau_\Pi^0)}6\tau^{-8/3}\,,
\end{equation}
leading to the identifications
\begin{equation}
\epsilon_0=\frac 32 \frac{N^2}{4\pi^2}\omega^4\,,\quad M_1=\eta_0\,, \quad M_2=\lambda_1^0-\eta_0 \tau_\Pi^0\,.
\end{equation}
These constants are in turn fixed by imposing regularity of the function $c(t,r)$ \cite{jp}. An expansion of this function around $v=w$ has a simple pole, unless we set
\begin{equation}
M_1=\frac 1{3w}\,, \qquad M_2=\frac{\ln2-1}{6 w^2} \label{eta0Sol}\,.
\end{equation}
Finally, the hydrodynamic coefficients are obtained by the relations
\begin{equation}
\eta=\epsilon_0\eta_0 \left(\frac{\epsilon}{\epsilon_0}\right)^{3/4}\,, \quad
\tau_\Pi=\tau_\Pi^0 \left(\frac{\epsilon}{\epsilon_0}\right)^{-1/4}\,,\quad
\lambda_1=\epsilon_0\lambda_1^0 \left(\frac{\epsilon}{\epsilon_0}\right)^{1/2} \label{ETL}\,.
\end{equation}

We now want to compute $\gamma$ corrections to these by using the $\gamma$ corrected action (\ref{5DAction}). Since it is not practical to find the equations of motions from the action (\ref{5DAction}), we will work in an effective action framework, as in \cite{beta}. To do this, first notice that in the metric (\ref{Metric}) there are two implicit constraints, namely
\begin{equation}
g_{rr}=0\,, \qquad g_{\tau r}=1\,.
\end{equation}
It is not correct to impose these constraints at the level of the action. They should only be imposed on the equations of motion. Therefore we modify the metric (\ref{Metric}) to:
\begin{eqnarray}
ds^2&=& g_{\mu\nu}dx^{\mu}dx^{\nu}\nonumber \\
&=& -r^2 a(t,r)d \tau^2+ 2\left(1+h(t,r)-\frac 12 a(t,r)g(t,r)\right) d \tau dr+\frac{g(t,r)}{r^2} dr^2\nonumber \\
&&+ r^2 \tau^2 e^{2b(t,r)-2c(t,r)}\left(1+\frac 1{r t}\right)^2 dy^2+r^2 e^c(dx_1^2+dx_2^2)\label{FullMetric}\,.
\end{eqnarray}
We further substitute
\begin{eqnarray}
a(t,r)&=& a_0(v)+u a_1(v)+u^2 a_2(v)+\gamma \hat a(t,r)\,, \nonumber \\
b(t,r)&=& b_0(v)+u b_1(v)+u^2 b_2(v)+\gamma \hat b(t,r)\,, \nonumber \\
c(t,r)&=& c_0(v)+u c_1(v)+u^2 c_2(v)+\gamma \hat c(t,r\,, \nonumber
\end{eqnarray}
and evaluate the action (\ref{5DAction}) on the modified metric. 
A few comments are in order:
\begin{itemize}
\item One only needs to evaluate the action to linear order in $g, h$.
\item It is sufficient to compute $W$ to linear order with respect to each and every field $\hat a,\hat b, \hat c, g, h$, meaning no mixed terms such as $\hat a \hat b$, $g \hat c$ can appear in $W$.
\item Since we are interested in computing $\hat a, \hat b, \hat c$ to quadratic order in $u$, it is sufficient to evaluate $W$ to linear order in $a_2,b_2,c_2$ and quadratic order in $a_1,b_1,c_1$.
\end{itemize}

Variation of the action $S=S(\hat a,\hat b, \hat c,g,h)$ with respect to the various fields lead to the equations of motion
\begin{equation}
\frac{\delta S}{\delta \hat a(t,r)}_{|g,h=0}=0\,, \quad
\frac{\delta S}{\delta \hat b(t,r)}_{|g,h=0}=0\,, \quad
\frac{\delta S}{\delta \hat c(t,r)}_{|g,h=0}=0\,,
\end{equation}
and to the constraints
\begin{equation}
\frac{\delta S}{\delta g(t,r)}_{|g,h=0}=0\,, \quad
\frac{\delta S}{\delta h(t,r)}_{|g,h=0}=0\,.
\end{equation}
After finding these equations, one expands $\hat a,\hat b, \hat c$ to quadratic order in $u$, exactly as it was done for the unhatted quantities. This leads to a set of equations for the hatted quantities $\hat a_i, \hat b_i, \hat c_i$, $i=0,1,2$.

\subsection{Order - 0}

It will be convenient to perform the change of variable $y\equiv w/v$. In terms of this coordinate the equations of motion are
\begin{eqnarray}
\hat b_0'' &=&S_B^0\,, \nonumber \\
\left(\hat c_0''-\frac 23 \hat b_0''\right)-\frac{3+y^4}{y(1-y^4)}\left(\hat c_0'-\frac 23 \hat b_0'\right)&=&S_C^0\nonumber\,, \\
\hat a_0''-\frac{6 \hat a_0'}{y}+\frac{12 \hat a_0}{y^2}-\frac{2 \left(y^4+3\right)
   \hat c_0'}{y}+2 (1-y^4) \hat c_0''&=&S_A^0 \label{EOMo0}\,.
\end{eqnarray}
The solution of this system of equations is straightforward. Imposing the boundary conditions $\hat a_0(0)=\hat b_0(0)=\hat c_0(0)=0$ leads to
\begin{eqnarray}
a_0(y)&=& A_1^0 y^4+A_2^0 y^3+\frac{15}{11} y^{12} \left(72-49 y^4\right)-\frac{2}{3} B_1^0y \left(1+y^4\right) \nonumber\,, \\
b_0(y)&=& -\frac{360}{11} y^{12}-B_1^0 y \nonumber \,,\\
c_0(y)&=& -\frac 23 B_1^0 y-\frac{240}{11}y^{12}+C_1^0 \ln(1-y^4)\,.
\end{eqnarray}
The coefficient $B_1^0$ is a gauge parameter \cite{jp} and we will set it to zero in what follows.
The constraint equations further impose $A_2^0=C_1^0=0$. Quite generally, the constraint equations can only affect the coefficients of the solutions to the homogeneous equations, and since these come from the supergravity part of the action, $C^4$ should not alter the constraints from those at the supergravity level. Another way to see this is to note 
that one can always expand the equations of motion about $y=0$, where all the source terms are negligible. The constraint equations must necessarily be the same around $y=0$ as at any other point, as they only fix constants. 

Expanding $a_0(y)$ around $y=0$ and taking the coefficient of $y^4$ we see that this solution modifies $\epsilon_0$ at order $\gamma$ by some undetermined constant $A_1^0$, which will however not affect any of the physical results as we will see.

\subsection{Order - 1}

The equations of motion are exactly the same as before (\ref{EOMo0}), performing the replacement $0\to 1$ in the indices of functions and sources. The solutions are now:

\begin{equation}
\begin{split}\nonumber
w~\! \hat a_1(y)=&A_1^1 y^4+A_2^1 y^3  -\frac{10 \left(-343 y^5-196 y^4+360 y+216\right) y^{12}}{11}\\
&-\frac {2(A_0^1 y^5+B_1^1(1+y^4)y)}{3}  \nonumber \,,\\
\end{split}
\end{equation}
\begin{equation}
\begin{split}\nonumber
w~\! \hat b_1(y)=&-B_1^1 y+\frac{720 y^{12} (2 y+1)}{11} \nonumber \,,\\
\end{split}
\end{equation}
\begin{equation}
\begin{split}
w~\! \hat c_1(y)=&
\frac{960 y^{13}}{11}+\frac{2873 y^{12}}{66}-\frac{490
   y^9}{99}+\frac{335 y^8}{22}-\frac{52 y^5}{11} +\frac{555
   y^4}{22}\\
   & +\frac{1}{792} (-528 B_1^1-2160) y+\frac{1}{792}
   \left(66 (A_1^0+15) \arctan \left(\frac{1}{y}\right)-33
   (A_1^0+15) \pi \right) \\
   & +\frac{(11 A_1^0+345) y \left(y^2+y+1\right)}{66
   \left(y^3+y^2+y+1\right)}  \\
& -\frac{\left(15+A_1^0\right) \mbox{arctanh}(y)+\left(2
   A_1^0-3\left(80+C_1^1\right)\right) \ln
   \left(1-y^4\right)}{12} \,.
\end{split}
\end{equation}
The constraint equations now imply $A_2^1=A_1^1+C_1^1=0$. We may further set $B_1^1$ to zero as we did at order zero. The constant $C_1^1$ is determined by imposing regularity at $y=1$ (that is, $v=w$). The function $\hat c_1(y)$ has a pole there unless we set $C_1^1=(A_1^0-165)/2$. 

Expanding $a_1(y)$ around $y=0$ and taking the coefficient of $y^4$ we get a non-zero contribution. This implies that to this order the energy density is now given by
\begin{equation}
\epsilon(\tau)=\frac 32 \frac{N^2}{4\pi^2}w^4 \left((1-A_1^0 \gamma)\tau^{-4/3}-2 \left(M_1+\gamma \frac{A_1^0-165}{4 w}\right) \tau^2\right)\,,
\end{equation}
with $M_1$ given by (\ref{eta0Sol}). Now, in the late time regime we assume that the expressions for the energy and entropy densities are given by
\begin{equation}
\epsilon=\frac 38 \pi^2 N^2 T^4(1+15\gamma)\,, \qquad s=\frac{\pi^2}2 N^2 T^3(1+15\gamma)\,.
\end{equation}
Then, this result together with (\ref{ETL}) leads to
\begin{equation}
\frac{\eta}{s}=\frac{1}{4\pi}(1+120 \gamma)\,,
\end{equation}
in perfect agreement with what was previously known in the literature \cite{bls,bb,bmis,beta}. Notice the undetermined constant $A_1^0$ has canceled out of the result.

\subsection{Order - 2}

The equations of motion are as in the previous sections, but now using the source terms of order 2. These are too long to display here, but can be obtained from the authors at request. For small $y$ all source terms vanish and so we get only the homogeneous equations solutions, namely
\begin{eqnarray}
w^2\hat a_2(y)=A_1^2 y^4+A_2^2 y^3 \nonumber\,, \\
w^2 b_2(y)=-B_1^2 y \nonumber \,,\\
w^3 c_2'(y)=C_1^2 y^5 \,.
\end{eqnarray}
One may once again choose $B_1^2=0$. The constraint equations force $A_2^2=2 A_1^2+C_1^2=0$. Therefore to determine $A_1^2$, which is the parameter that enters the energy density expansion, we must find out the value of $C_1^2$. This can be determined by imposing regularity of the full solution at $y=1$. Close to $y=1$ one has
\begin{equation}
c_2'(y)=\frac{(1+2 \ln 2) A_1^0+18 C_1^2-510 \ln 2-1085}{72
   (1-y)}+(\mbox{Regular at $y=1$})\,.
\end{equation}
Imposing regularity forces us to pick
\begin{equation}
C_1^2=\frac{1085+510 \ln 2-A_1^0 (1+\ln 4)}{18}=-2 A_1^2\,.
\end{equation}
The energy density expansion coefficient of $\tau^{-8/3}$ receives a modification, namely
\begin{equation}
\frac{9 \eta_0^2+4(\lambda_1^0-\eta_0\tau_\Pi^0)}6=\frac{9 M_1^2+4 M_2}6-\gamma A_1^2\,. 
\end{equation}
Then, using (\ref{ETL}) and the known result \cite{bp}
\begin{equation}
\tau_\Pi T=\frac{2-\ln 2}{2\pi}+\frac{375}{4\pi}\ga\,,
\end{equation}
together with the expression for $\eta/s$ found in the previous section, leads to
\begin{equation}
\frac{\lambda_1 T}{\eta}=\frac 1{2\pi}(1+215 \gamma)\,,
\end{equation}
in agreement with \eqref{lambda1}.

\section*{Acknowledgments}
We would like to thank Shin Nakamura for valuable comments. 
Research at Perimeter Institute is supported by the Government of
Canada through Industry Canada and by the Province of Ontario
through the Ministry of Research \& Innovation.
AB gratefully acknowledges further support by an NSERC Discovery
grant and support through the Early Researcher Award program by the
Province of Ontario. 
MP is supported by the Portuguese Fundacao para a Ciencia e Tecnologia, grant SFRH/BD/23438/2005.

\appendix
\section{$\{a_2,b_2,c_2\}$ and $a_3$}\label{background}

Explicit analytic solutions for $\{a_2,b_2,c_2\}$ were found in \cite{hj}:
\begin{equation}
\begin{split}
a_2=&\frac{(9+5v^4)v^2}{12(9-v^8)} -C \frac{(9+v^4)v^4}{72(9-v^8)} +
\he^2 \frac{(-1053-171v^4+9v^8+7v^{12})v^4}{6(9-v^8)^2}\\
&+
\frac{1}{8\sqrt{3}} \ln \frac{\sqrt{3}-v^2}{\sqrt{3}+v^2}-
\frac{3}{4} \he^2 \ln \frac{3-v^4}{3+v^4}\,,  \\
c_2 =& -\frac{\pi^2}{288\sqrt{3}} +\frac{v^2(9+v^4)}{12(9-v^8)} +C
\frac{v^4}{72(3+v^4)} -\he^2 \frac{(-9+54v^4+7v^8)v^4}{6(3+v^4)(9-v^8)}\\
& +
\frac{1}{8\sqrt{3}} \ln \frac{\sqrt{3}-v^2}{\sqrt{3}+v^2}+\frac{1}{72} (C+66\he^2) \ln \frac{3-v^4}{3+v^4}\\
&+
 \frac{1}{24\sqrt{3}} \left( \ln \frac{\sqrt{3}-v^2}{\sqrt{3}+v^2}
\ln\frac{(\sqrt{3}-v^2)(\sqrt{3}+v^2)^3}{4(3+v^4)^2} -{\rm li}_2 \left(-
\frac{(\sqrt{3}-v^2)^2}{(\sqrt{3}+v^2)^2} \right)\right)\,,\\
b_2 =& -2 c_2+\frac{v^2}{4(3+v^4)} +C\frac{v^4}{24(3+v^4)} +\he^2
\frac{(39+7v^4)v^4}{2(3+v^4)^2} +
\frac{1}{8\sqrt{3}} \ln \frac{\sqrt{3}-v^2}{\sqrt{3}+v^2}\\
&+
\frac{3}{4} \he^2 \ln \frac{3-v^4}{3+v^4}\,,
\end{split}
\eqlabel{order2}
\end{equation}
where $\{\he,C\}$ are arbitrary parameters.

At third order (for the warp factors $\{a_3,b_3,c_3\}$) one obtains four second order ODE's and a first order constraint, 
all linear in $\{a_3,b_3,c_3\}$ (and their derivatives). The two additional equations are the constraints used to fix the radial coordinate and the 
late time in the boost-invariant metric ansatz \eqref{5dimM}. It is straightforward to use these constraints to solve algebraically for 
$b_3'$ and $c_3'$ in terms of $a_3'$ and the lower order warp factors. These expressions can further be used to obtain a decoupled 
second order ODE for $a_3$: 
\begin{equation}
\begin{split}
0=a_3''+\frac{5 v^{16}+18 v^{12}+216 v^8+126 v^4+243}{(v^8-9) v (v^4+2 v^2+3) (v^4-2 v^2+3)} a_3'+\calj_{a_3}\,,
\end{split}
\eqlabel{order3}
\end{equation} 
\begin{equation}
\begin{split}
\calj_{a_3}=&\frac{2\he (5 v^{12}+9 v^8+27 v^4+135)}{3(v^4-3)^2 (v^4+2 v^2+3) (v^4-2 v^2+3)}  \ln\frac{3-v^4}{3+v^4}
\\
&+\frac{32 \sqrt{3} v^6 \he (v^8+18 v^4+9)}{(v^8-9)^2 (v^4+2 v^2+3) (v^4-2 v^2+3)}  \arctanh\frac{v^2}{\sqrt{3}}
\\
&-\frac{4\he v^4}{3(v^4-2 v^2+3) (v^4+2 v^2+3) (v^8-9)^5}  \biggl(295245+393660 v^4-531 v^{32}\\
&-2556 v^{28}-9882 v^{24}-21708 v^{20}
+27702 v^{16}+358668 v^{12}+898857 v^8+v^{40}\\
&-42768 \he^2 v^{26}-1113264 v^{22} \he^2-2554416 \he^2 v^{18}-7336656 v^{14} \he^2
-12912048 \he^2 v^{10}\\
&-46294416 \he^2 v^6-9447840 \he^2 v^2+432 v^{30} C+12 v^{34} C-52488 v^{10} C+27216 v^{14} C\\
&-78732 v^2 C-3024 v^{22} C
+1296 v^{30} \he^2+144 v^{34} \he^2-314928 v^6 C+648 v^{26} C\biggr)\,.
\end{split}
\eqlabel{order31}
\end{equation} 
For the computation of $C$ we actually need only the asymptotic solution of $a_3'(v)$ as $x\equiv 3^{1/4}-v\to 0_+$.
Using \eqref{order3} and \eqref{order31} we find:
\begin{equation}
\begin{split}
\frac{da_3(v)}{dv}\bigg|_{x\equiv 3^{1/4}-v\to 0_+}=&\frac{\he^3 3^{3/4}}{8}\ \frac {1}{x^{4}}+
\left(\frac{11\he^3 \sqrt3}{24} -\frac{\he}{12} +\frac{\sqrt 3\he C}{144}\right)\ \frac{1}{x^3}+\cala_3\ \frac{1}{x^2}+\calo(x^{-1})\,,
\end{split}
\eqlabel{expa3}
\end{equation}
where $\cala_3$ must be fixed so it satisfies the boundary condition \eqref{bbcc}.

\section{Data for computing $\cali^{[2]W}_3$}\label{riem2}

The coefficients $\calq_i$ in \eqref{i3w} are given by:
\begin{equation}
\begin{split}
&\calq_1=102036672+177147 \delta_1+(-177147 \delta_1-102036672) v^4+(-452709 \delta_1\\
&-8039875644) v^8+(-229635 \delta_1+17670478860) v^{12}+(4374 \delta_1-10087292556) v^{16}\\
&+(2301024348+42282 \delta_1) v^{20}+(14094 \delta_1-758889972) v^{24}+(373696740+162 \delta_1) v^{28}\\
&-(945 \delta_1+73262340) v^{32}+(-207 \delta_1+3556980) v^{36}-9 v^{40} \delta_1+v^{44} \delta_1\,,
\end{split}
\eqlabel{cq1}
\end{equation}
\begin{equation}
\begin{split}
&\calq_2=-258280326 \he_0 \delta_1-74384733888 \he_0+43046721 \delta_2+(-86093442 \he_0 \delta_1\\
&+41324852160 \he_0+52612659 \delta_2) v^4+(-1004423490 \he_0 \delta_1+9228872608632 \he_0\\
&+113196933 \delta_2) v^8+(158192271 \delta_2-1491223446 \he_0 \delta_1-33853663263240 \he_0) v^{12}\\
&+(105323733 \delta_2-798224382 \he_0 \delta_1+69144574698756 \he_0) v^{16}-(38887909614036 \he_0\\
&-32483511 \delta_2+170809074 \he_0 \delta_1) v^{20}-(14604786 \he_0 \delta_1273375 \delta_2\\
&-22636564766388 \he_0) v^{24}-(4704237 \delta_2+12251574 \he_0 \delta_1+5034336682644 \he_0) v^{28}\\
&-(2229525 \delta_2+5466042 \he_0 \delta_1-4161674125224 \he_0) v^{32}-(568215 \delta_2+906940950984 \he_0\\
&-1465290 \he_0 \delta_1) v^{36}+(371486277216 \he_0-76545 \delta_2+1625994 \he_0 \delta_1) v^{40}-(49680556848 \he_0\\
&-6813 \delta_2-451278 \he_0 \delta_1) v^{44}+(47142 \he_0 \delta_1+18898470612 \he_0+9399 \delta_2) v^{48}+(3821 \delta_2\\
&-1110 \he_0 \delta_1-2382365412 \he_0) v^{52}+(-342 \he_0 \delta_1+923 \delta_2+258200244 \he_0) v^{56}+(129 \delta_2\\
&-4642164 \he_0+126 \he_0 \delta_1) v^{60}+(8 \delta_2+24 \he_0 \delta_1) v^{64}\,,
\end{split}
\eqlabel{cq2}
\end{equation}
\begin{equation}
\begin{split}
&\calq_3=-387420489 \he_0 \delta_1-148769467776 \he_0+43046721 \delta_2+(-1607077584 \he_0 \delta_1\\
&+153055008 \delta_2-661197634560 \he_0) v^4+(2210707403676 \he_0-3386342052 \he_0 \delta_1\\
&+267846264 \delta_2) v^8+(10723033860480 \he_0+232416864 \delta_2-3409725456 \he_0 \delta_1) v^{12}
\\
&+(-2092224168 \he_0 \delta_1+103060188 \delta_2+18352430617176 \he_0) v^{16}
+(18685728 \delta_2\\
&-6136825576320 \he_0-970923024 \he_0 \delta_1) v^{20}
+(15039362941668 \he_0-4846392 \delta_2\\
&-352115748 \he_0 \delta_1) v^{24}
+(-1235549044224 \he_0-5155488 \delta_2-74427984 \he_0 \delta_1) v^{28}
\\
&+(2004210714960 \he_0+6647022 \he_0 \delta_1-2215674 \delta_2) v^{32}
+(11706768 \he_0 \delta_1\\
&-131529982464 \he_0-572832 \delta_2) v^{36}
+(4706100 \he_0 \delta_1+188278471332 \he_0-59832 \delta_2) v^{40}
\\
&+(25632 \delta_2-7695285120 \he_0+1178064 \he_0 \delta_1) v^{44}
+(15708 \delta_2+224640 \he_0 \delta_1\\
&+2953736280 \he_0) v^{48}
+(34128 \he_0 \delta_1+3936 \delta_2+208054656 \he_0) v^{52}
+(3348 \he_0 \delta_1+504 \delta_2\\
&+6959196 \he_0) v^{56}
+(144 \he_0 \delta_1+32 \delta_2) v^{60}
+(\delta_2+3 \he_0 \delta_1) v^{64}\,,
\end{split}
\eqlabel{cq3}
\end{equation}
\begin{equation}
\begin{split}
&\calq_4=(6025163444928 \he_0^3+69735688020 \he_0^3 \delta_1-13947137604 \he_0^2 \delta_2) v^2-278942752080 \he_0\\
&+(539289320688 \he_0^3 \delta_1-86782189536 \he_0^2 \delta_2-301239576062928 \he_0^3) v^{10}
-645700815 \delta_2\\
&+3874204890 \he_0 \delta_1
+(378122397264 \he_0-1822311189 \delta_2+12957063021 \he_0 \delta_1) v^4
\\
&+(34044733308852 \he_0-2368101096 \delta_2+29635275924 \he_0 \delta_1) v^{12}
+(-163296 \he_0^2 \delta_2\\
&-869166288 \he_0^3-34992 \he_0^3 \delta_1) v^{58}
+(518856477120 \he_0^3 \delta_1-75303063936 \he_0^2 \delta_2\\
&-1212305862965760 \he_0^3) v^{14}
+(6585380676 \he_0 \delta_1-63131883819804 \he_0-83613384 \delta_2) v^{24}
\\
&+(-1405753115292 \he_0-73136196 \he_0 \delta_1+2163672 \delta_2) v^{40}
+(-7464960 \he_0^3+5184 \he_0^3 \delta_1\\
&-10368 \he_0^2 \delta_2) v^{62}
-(6931186200 \he_0+76284 \delta_2+1076544 \he_0 \delta_1) v^{52}
-(49589822592 \he_0^2 \delta_2\\
&-272744024256 \he_0^3 \delta_1-35704672266240 \he_0^3) v^6
+(-1501497972 \delta_2-122924025168840 \he_0\\
&+25694109468 \he_0 \delta_1) v^{16}
+(4697466048 \he_0^3 \delta_1+1670378112 \he_0^2 \delta_2+134641969016832 \he_0^3)\\
&\times v^{30}
-(5 \delta_2+15 \he_0 \delta_1+15120 \he_0) v^{68}
-(1275264 \he_0^2 \delta_2+23388092928 \he_0^3+1135296 \he_0^3 \delta_1)\\
&\times v^{54}
+(-5089392 \he_0^2 \delta_2-322667880480 \he_0^3-14082336 \he_0^3 \delta_1) v^{50}
+(292743211968 \he_0^3 \delta_1\\
&-33391500912 \he_0^2 \delta_2-2006104387311648 \he_0^3) v^{18}
+(-127 \delta_2-384048 \he_0-522 \he_0 \delta_1) v^{64}
\\&
+(-58217292 \he_0-10116 \he_0 \delta_1-1608 \delta_2) v^{60}
+(-6054175872 \he_0^2 \delta_2+116968038336 \he_0^3 \delta_1\\
&+658418155614720 \he_0^3) v^{22}
+(44654166 \delta_2+183812976 \he_0 \delta_1-16291341648432 \he_0) v^{32}
\\
&+(15604524936 \he_0 \delta_1-596867292 \delta_2-119986465909464 \he_0) v^{20}
+(19385568 \he_0^2 \delta_2\\
&-20320117294896 \he_0^3-547029936 \he_0^3 \delta_1) v^{42}
+(185597568 \he_0^2 \delta_2-1635526080 \he_0^3 \delta_1\\
&+14338868355072 \he_0^3) v^{38}
+(-6316380 \he_0 \delta_1-272916 \delta_2-58737530376 \he_0) v^{48}
\\
&-(26068068 \he_0 \delta_1+275114490228 \he_0+344088 \delta_2) v^{44}
-(110621376 \he_0^3 \delta_1+8304768 \he_0^2 \delta_2\\
&-825111360000 \he_0^3) v^{46}
+(-36868394133108 \he_0+1799419860 \he_0 \delta_1+58419144 \delta_2) v^{28}
\\
&+(14884722 \delta_2-105925158 \he_0 \delta_1-5478599707344 \he_0) v^{36}
+(-2153635128 \he_0^3 \delta_1\\
&+717878376 \he_0^2 \delta_2-215937884784960 \he_0^3) v^{34}
+(34888038768 \he_0^3 \delta_1\\
&-1623120631374384 \he_0^3+1570231008 \he_0^2 \delta_2) v^{26}
+(-233280 \he_0^3-324 \he_0^2 \delta_2+324 \he_0^3 \delta_1) v^{66}
\\
&+(107373276 \he_0-13368 \delta_2-127188 \he_0 \delta_1) v^{56}
+(-2563671384 \delta_2+23819185620 \he_0 \delta_1\\
&-81312328791972 \he_0) v^8\,,
\end{split}
\eqlabel{cq4}
\end{equation}
\begin{equation}
\begin{split}
&\calq_5=5782609521 v^4+3314566899 v^{24}+6632626437 v^{20}+10567349991 v^{16}
\\
&+12018183921 v^{12}+9093486951 v^8+1375999893 v^{28}+26344593252 \he_0^2 v^2\\
&-146014477632 \he_0^2 v^6+10005971148 \he_0^2 v^{10}
-59066478504 v^{14} \he_0^2-75754434492 \he_0^2 v^{18}\\
&-55948533840 v^{22} \he_0^2-22400959860 \he_0^2 v^{26}
+26907093 v^{40}+4639059 v^{44}+126080307 v^{36}\\
&+467481699 v^{32}-25 v^{60}+624309 v^{48}+58263 v^{52}+v^{64}+2961 v^{56}-2708022132 v^{34}
 \he_0^2\\
&-740454048 v^{38} \he_0^2-108490428 v^{42} \he_0^2-16471512 v^{46} \he_0^2-662580 v^{50} \he_0^2\\
&-9226708056 v^{30} \he_0^2-92880 v^{54} \he_0^2+216 v^{62} \he_0^2+2916 v^{58} \he_0^2+2754990144\,,
\end{split}
\eqlabel{cq5}
\end{equation}
\begin{equation}
\begin{split}
&\calq_6=-12223143-20785248 v^{4}-2196720 v^{24}-6722352 v^{20}
-15363675 v^{16}\\
&-23024736 v^{12}-23225940 v^8
-579312 v^{28}-172186884 \he_0^2 v^2+752520456 \he_0^2 v^6\\
&-108413964 \he_0^2 v^{10}
+166754376 v^{14} \he_0^2+257348664 \he_0^2 v^{18}+140702832 v^{22} \he_0^2\\
&+37144008 \he_0^2 v^{26}-828 v^{40}+16 v^{44}
-14640 v^{36}-114813 v^{32}-v^{48}+2349324 v^{34} \he_0^2\\
&+425736 v^{38} \he_0^2+15876 v^{42} \he_0^2+648 v^{46} \he_0^2
+12227760 v^{30} \he_0^2\,,
\end{split}
\eqlabel{cq6}
\end{equation}
\begin{equation}
\begin{split}
&\calq_7=163046532875760 v^{32}-108446607008910 v^{36}+26922194450352 v^{40}\\
&-11404592789688 v^{44}+3756782036502 v^{48}
-421486536163500 v^{28}\\
&-1062657813600 v^{52}-45397293396 v^{60}+290922851196 v^{56}-1976417606251692 v^{20}\\
&+2525883800478795 v^{16}-2045958218223822 v^{12}-68193367053291 v^4\\
&+518822412814782 v^8
-1129718145924+703601320689252 v^{24}+20 v^{84}-2958074 v^{76}\\
&-1437073423 v^{68}+13224991788 v^{64}+175580978 v^{72}+140423 v^{80}\,,
\end{split}
\eqlabel{cq7}
\end{equation}
\begin{equation}
\begin{split}
&\calq_8=8 v^{80}-412 v^76+1130201 v^{72}-36420582 v^{68}+661666116 v^{64}
-4050893934 v^{60}\\
&+31716802512 v^{56}-92949815898 v^{52}+550493727984 v^{48}
-1678881463974 v^{44}\\
&+4547507588514 v^{40}-13314555136710 v^{36}+9923931144288 v^{32}
-82854890580186 v^{28}\\
&+1492061941800 v^{24}
-51956321216238 v^{20}-427117486358664 v^{16}\\
&+308048735088522 v^{12}
-243317414920563 v^8+4198088418804 v^4+1938652126956\,.
\end{split}
\eqlabel{cq8}
\end{equation}

In \cite{beta} an expression for $f_2'\equiv \hc_2'+\frac 12\hb_2'$ was presented. It is straightforward to 
integrate it to obtain (imposing the proper boundary condition, \ie, $f_2(v)\to 0 $ as $v\to 0_+$):
\begin{equation}
\begin{split}
&f_2=\frac{\sqrt 3(\delta_1-954)}{288} \arctanh\frac{v^2}{\sqrt 3}+\frac{3475\sqrt 3 (v^4-3)}{128(3+v^4)}\arctan\frac{v^2}{\sqrt 3}
\\
&-\frac{\he_0 (1080 \he_0+\delta_2-3 \he_0 \delta_1)}{24} \ln\frac{3-v^4}{3+v^4}+\frac{v^2}{1152 (3+v^4)^4 (v^8-9)^4} \biggl(-2125764 \delta_1
\\
&+(1417176 \delta_1 C_0+272097792 \he_0^2 \delta_1+226521411840 \he_0^2-51018336 \he_0 \delta_2+408146688 C_0\\
&+17006112 \delta_3) v^2
+(1180980 \delta_1 C_0+272097792 C_0+158723712 \he_0^2 \delta_1-39680928 \he_0 \delta_2\\
&+17006112 \delta_3+174686782464 \he_0^2) v^6
+(11074698999-4960116 \delta_1) v^4
+(-8468512335\\
&-2598156 \delta_1) v^8
+(-314928 \delta_1 C_0-100147104 \he_0^2 \delta_1-2857026816 C_0+13226976 \he_0 \delta_2\\
&-816701522688 \he_0^2-1889568 \delta_3) v^{10}
+(3564 \delta_1+91909647) v^{32}
+(-165888 \he_0^2 \delta_1\\
&-597756672 \he_0^2+3732480 C_0-1512 \delta_1 C_0+44064 \he_0 \delta_2-28512 \delta_3) v^{34}
+(866052 \delta_1\\
&+22334044221) v^{12}
+(-74952864 \he_0^2 \delta_1-498636 \delta_1 C_0+3355872768 C_0+17006112 \he_0 \delta_2\\
&+1203035037696 \he_0^2-6928416 \delta_3) v^{14}
+(6110429886+997272 \delta_1) v^{16}
+(419904 \he_0 \delta_2\\
&+8398080 \he_0^2 \delta_1-1375605504 C_0-1259712 \delta_3-34992 \delta_1 C_0-1340988618240 \he_0^2) v^{18}
\\
&+(979776 \delta_3+481310760960 \he_0^2+75816 \delta_1 C_0+10077696 C_0-2659392 \he_0 \delta_2\\
&+12877056 \he_0^2 \delta_1) v^{22}
+(87480 \delta_1+9442000710) v^{20}+(6220800 \he_0^2-2592 \he_0 \delta_2\\
&+17280 \he_0^2 \delta_1+36 \delta_1 C_0-864 \delta_3) v^{38}
+(-3125709+1404 \delta_1) v^{36}
+(-1440 \he_0 \delta_2+864 \delta_3\\
&+6324480 \he_0^2+48 \delta_1 C_0+6624 \he_0^2 \delta_1) v^{42}+(36 \delta_1+645597) v^{40}
+(-12 \delta_1+105273) v^{44}\\
&+(480 \he_0^2 \delta_1+96 \delta_3+4 \delta_1 C_0-96 \he_0 \delta_2+552960 \he_0^2) v^{46}
+(-110808 \delta_1+3107022786) v^{24}\\
&+(-121224605184 \he_0^2-419904 \he_0 \delta_2+153964800 C_0+15552 \delta_1 C_0+886464 \he_0^2 \delta_1\\
&+326592 \delta_3) v^{26}
+(-25272 \delta_1+208396314) v^{28}+(171072 \he_0 \delta_2-917568 \he_0^2 \delta_1\\
&-42923520 C_0+7936745472 \he_0^2-4536 \delta_1 C_0-46656 \delta_3) v^{30}+18648796131\biggr)\,.
\end{split}
\eqlabel{f2res}
\end{equation}
We can further use a constraint at the second order (similar to \eqref{eom35}) to find
\begin{equation}
\begin{split}
&\ha_2'=-\frac{10425v^3 (v^8+2 v^4+9)\sqrt 3}{8(v^8-9)^2}   \arctan\frac{v^2}{\sqrt 3}
+\frac{v^3}{72 (v^8-9)^5 (3+v^4)^4} \biggl(76527504 \delta_3\\
&+1836660096 C_0+1101996057600 \he_0^2+994857552 \he_0^2 \delta_1-153055008 \he_0 \delta_2\\
&+6377292 \delta_1 C_0
+(-864 \he_0 \delta_2+12 \delta_1 C_0+576 \he_0^2 \delta_1-2016 \delta_3) v^{44}
+(119042784 \delta_3\\
&-289103904 \he_0 \delta_2+1851353376768 \he_0^2+1938696768 \he_0^2 \delta_1+3265173504 C_0\\
&+10628820 \delta_1 C_0) v^4
+(-339560128512 \he_0^2-3129840 \he_0^2 \delta_1+674179200 C_0\\
&+15228 \delta_1 C_0+190512 \delta_3-738720 \he_0 \delta_2) v^{32}
+(1912896 \he_0 \delta_2+2034579326976 \he_0^2\\
&+606528 \delta_3-21384 \delta_1 C_0-22581504 \he_0^2 \delta_1-2058089472 C_0) v^{28}
+(96435034542\\
&+1749600 \delta_1) v^{22}
+(194755671-17712 \delta_1) v^{38}
-(24091992 \delta_1-445385644605) v^{10}\\
&+(24383227336704 \he_0^2+69838433280 C_0-5668704 \delta_3+2913084 \delta_1 C_0\\
&-179508960 \he_0 \delta_2+1795089600 \he_0^2 \delta_1) v^{12}
+(-25509168 \delta_1-86716822293) v^6\\
&-144 \delta_3 v^{48}
+(18828186993-12282192 \delta_1) v^{14}+(-9977140749312 \he_0^2\\
&-78941952 C_0-71313696 \he_0^2 \delta_1+16096320 \he_0 \delta_2-274104 \delta_1 C_0) v^{24}
+(5598720 C_0\\
&-8640 \delta_3+468 \delta_1 C_0-4752 \he_0^2 \delta_1-28512 \he_0 \delta_2-618098688 \he_0^2) v^{40}
+(314464150422\\
&-1679616 \delta_1) v^{18}
+(-27864 \delta_1+41837067) v^34-(57209092209+12754584 \delta_1) v^2\\
&+(178848 \delta_1+11908840482) v^{30}
+(21413128771584 \he_0^2-3359232 \he_0^2 \delta_1-5458752 \delta_3\\
&+18280940544 C_0-647352 \delta_1 C_0+28133568 \he_0 \delta_2) v^{20}
+(7776 \delta_3-212544 \he_0^2 \delta_1\\
&-261792 \he_0 \delta_2-93312000 C_0+16858865664 \he_0^2+4644 \delta_1 C_0) v^{36}+(1038096 \delta_1\\
&+32981460282) v^{26}
+(7794468 \delta_1 C_0-33604077312 C_0+2406364848 \he_0^2 \delta_1\\
&-8771888618496 \he_0^2-300441312 \he_0 \delta_2+56687040 \delta_3) v^8
+(-336 \delta_1-357507) v^{46}\\
&+(-16 \delta_1-4608) v^{50}
+(-28973376 \he_0 \delta_2+619148448 \he_0^2 \delta_1-54268392960 C_0\\
&-15431472 \delta_3-36532030952448 \he_0^2-52488 \delta_1 C_0) v^{16}
+(-4094631-3384 \delta_1) v^{42}\biggr)\,.
\end{split}
\eqlabel{hares}
\end{equation}

Using \eqref{eom31}-\eqref{eom35} we can obtain a decoupled equation for $\ha_3$:
\begin{equation}
\begin{split}
0=\ha_3''+\frac{5 v^{16}+18 v^{12}+216 v^8+126 v^4+243}{(v^8-9) v (v^4+2 v^2+3) (v^4-2 v^2+3)} \ha_3'+\calj_{\ha_3}\,,
\end{split}
\eqlabel{haorder3}
\end{equation} 
\begin{equation}
\begin{split}
\calj_{\ha_3}=&-\frac{768 (v^8+18 v^4+9) v^6 \he_0}{(v^8-9)^2 (v^8+2 v^4+9)} f_2'-\frac{4 v^3\calr_1}{3 (v^4-2 v^2+3)^2 (v^4+2 v^2+3)^2 (v^8-9)^6 (3+v^4)}\\
&\times a_3'-\frac{16 v^6 \sqrt3\calr_2}{9 (v^8-9)^6 (3+v^4)^2 (v^4-2 v^2+3)^2 (v^4+2 v^2+3)^2}  \arctanh\frac{v^2}{\sqrt 3}\\
&-\frac{125100 v^6 \he_0 \sqrt3\calr_3}{(v^8-9)^4 (v^4-3)^2 (v^4-2 v^2+3) (v^4+2 v^2+3)} \arctan\frac{v^2}{\sqrt3}\\
&+\frac{\calr_4}{27 (v^8-9)^5 (v^4-3) (v^4-2 v^2+3)^2 (v^4+2 v^2+3)^2} \ln\frac{3-v^4}{3+v^4}\\
&+\frac{2 \he_0 \delta_1 v^4\calr_5}{9 (v^8-9)^6 (v^4-2 v^2+3)^2 (v^4+2 v^2+3)^2}\\
& +\frac{2 v^4 \delta_2\calr_6}{27 (v^8-9)^5 (v^4-3) (v^4-2 v^2+3) (v^4+2 v^2+3)}\\
&-\frac{192 v^6 \he_0 \delta_3\calr_7}{(v^8-9)^4 (v^4-3)^2 (v^4-2 v^2+3) (v^4+2 v^2+3)}\\
&+\frac{4 v^4 \he_0\calr_8}{(v^8-9)^7 (v^4+2 v^2+3)^2 (v^4-2 v^2+3)^2 (3+v^4)^4}\,,
\end{split}
\eqlabel{haorder31}
\end{equation} 
with
\begin{equation}
\begin{split}
&\calr_1=14348907 \delta_1+8264970432+(-1435349865024+90876411 \delta_1) v^4
+(46235367 \delta_1\\
&+6473308508352) v^8
+(-23206257 \delta_1-11722074916032) v^{12}
+(11877204669696\\
&-16592769 \delta_1) v^{16}
+(1043199 \delta_1-7723609200000) v^{20}
+(1646811 \delta_1\\
&+3567784040064) v^{24}
+(-85293 \delta_1-1321470640512) v^{28}
+(-28431 \delta_1\\
&+440473837248) v^{32}
+(-132105017664+60993 \delta_1) v^{36}
+(4293 \delta_1+31786872768) v^{40}
\\
&+(-5435190720-7587 \delta_1) v^{44}
+(594864000-1179 \delta_1) v^{48}
+(261 \delta_1-36391680) v^{52}
\\
&+(933120+57 \delta_1) v^{56}+v^{60} \delta_1\,,
\end{split}
\eqlabel{resr1}
\end{equation}
\begin{equation}
\begin{split}
&\calr_2=-16529940864 \he_0+4782969 \delta_2-43046721 \he_0 \delta_1+(13817466 \delta_2+532631427840 \he_0\\
&-149866362 \he_0 \delta_1) v^4
+(8680203 \delta_2+3526999604352 \he_0-101505231 \he_0 \delta_1) v^8
\\
&+(6608711172096 \he_0-1653372 \delta_2+7794468 \he_0 \delta_1) v^{12}
+(-3247695 \delta_2+26394903 \he_0 \delta_1\\
&-1430009945856 \he_0) v^{16}
+(5786802 \he_0 \delta_1+5096210245632 \he_0-800442 \delta_2) v^{20}
\\
&-(1358127 \he_0 \delta_1
+277988205312 \he_0-247131 \delta_2) v^{24}
+(726514541568 \he_0+180792 \delta_2\\
&-542376 \he_0 \delta_1) v^{28}
+(-30848107392 \he_0+27459 \delta_2-13851 \he_0 \delta_1) v^{32}
+(-9882 \delta_2\\
&+62892101376 \he_0-12150 \he_0 \delta_1) v^{36}
+(-1974761856 \he_0-4455 \delta_2-9477 \he_0 \delta_1) v^{40}
\\
&+(-252 \delta_2+324 \he_0 \delta_1+1007023104 \he_0) v^{44}
+(147 \delta_2+837 \he_0 \delta_1+60466176 \he_0) v^{48}
\\
&+(1119744 \he_0+126 \he_0 \delta_1+26 \delta_2) v^{52}
+(3 \he_0 \delta_1+\delta_2) v^{56}\,,
\end{split}
\eqlabel{resr2}
\end{equation}
\begin{equation}
\begin{split}
&\calr_3=7 v^{28}-3 v^{24}-93 v^{20}-63 v^{16}+1269 v^{12}-4617 v^8+6561 v^4+2187\,,
\end{split}
\eqlabel{resr3}
\end{equation}
\begin{equation}
\begin{split}
&\calr_4=-51656065200 \he_0-23914845 \delta_2+(459165024 \he_0^3 \delta_1-153055008 \he_0^2 \delta_2\\
&-165299408640 \he_0^3) v^6
+71744535 \he_0 \delta_1+(596465856 \he_0+684 \delta_2-2970 \he_0 \delta_1) v^{44}
\\
&+(-12397455648 \he_0+100442349 \he_0 \delta_1-18068994 \delta_2) v^4
+(1948617 \delta_2\\
&-3275759808720 \he_0+39326634 \he_0 \delta_1) v^8
+(-312232216320 \he_0^3+867311712 \he_0^3 \delta_1\\
&-289103904 \he_0^2 \delta_2) v^{10}
+(-87 \he_0 \delta_1-102816 \he_0-34 \delta_2) v^{52}
+(6022998 \he_0 \delta_1+4487724 \delta_2\\
&+3920759121600 \he_0) v^{12}
+(1539 \delta_2-3078 \he_0 \delta_1-13050184752 \he_0) v^{40}
+(28512 \he_0^2 \delta_2\\
&-85536 \he_0^3 \delta_1+30792960 \he_0^3) v^{42}
+(1259712 \he_0^2 \delta_2+1360488960 \he_0^3-3779136 \he_0^3 \delta_1) v^{22}
\\
&+(67344203520 \he_0^3-187067232 \he_0^3 \delta_1+62355744 \he_0^2 \delta_2) v^{14}
+(368145 \he_0 \delta_1-12879 \delta_2\\
&-282820705488 \he_0) v^{32}
+(411505920 \he_0^3+381024 \he_0^2 \delta_2-1143072 \he_0^3 \delta_1) v^{38}
-(115911 \delta_2\\
&+2547194665968 \he_0+2965572 \he_0 \delta_1) v^{24}
+(2592 \he_0^3 \delta_1-933120 \he_0^3-864 \he_0^2 \delta_2) v^{50}
-(5 \delta_2\\
&+15120 \he_0) v^{56}
+(-64152 \delta_2+96228 \he_0 \delta_1+143399735424 \he_0) v^{28}
+(50388480 \he_0^3\\
&+46656 \he_0^2 \delta_2-139968 \he_0^3 \delta_1) v^{34}
+(-765 \he_0 \delta_1+33 \delta_2-55887408 \he_0) v^{48}
+(44064 \he_0^3 \delta_1\\
&-15863040 \he_0^3-14688 \he_0^2 \delta_2) v^{46}
+(26453952 \he_0^3 \delta_1-9523422720 \he_0^3-8817984 \he_0^2 \delta_2) v^{26}
\\
&+(92588832 \he_0^2 \delta_2-277766496 \he_0^3 \delta_1+99995938560 \he_0^3) v^{18}
+(68283 \he_0 \delta_1+162 \delta_2\\
&+23803727904 \he_0) v^{36}
+(1924904811168 \he_0-5570289 \he_0 \delta_1+13122 \delta_2) v^{20}
\\
&+(8817984 \he_0^3 \delta_1-3174474240 \he_0^3-2939328 \he_0^2 \delta_2) v^{30}
+(-1121931 \he_0 \delta_1\\
&-9513261568080 \he_0+1121931 \delta_2) v^{16}\,,
\end{split}
\eqlabel{resr4}
\end{equation}
\begin{equation}
\begin{split}
&\calr_5=-(7164612 C_0+4472292528 \he_0^2) v^{18}
+119062467 v^{16}
+223087122 v^{12}\\
&+315675954 v^8+74933181 v^4-9198 v^{44}-1583469 v^{32}-82638 v^{40}
-430029 v^{36}\\
&-915 v^{48}+13 v^{52}-2032452 v^{28}+34031907 v^{20}+4452732 v^{24}
+(115560 C_0\\
&-16708032 \he_0^2) v^{34}
+(-46530612 C_0-10489936752 \he_0^2) v^{10}+(17712 C_0-3045600 \he_0^2) v^{38}
\\
&+(52704 \he_0^2+648 C_0) v^{46}
+(12 C_0+1872 \he_0^2) v^{50}
+(-118646208 \he_0^2+470448 C_0) v^{30}\\
&+(110160 \he_0^2+2268 C_0) v^{42}
+(-10507887648 \he_0^2-15903864 C_0) v^{14}+(-482772960 \he_0^2\\
&+173016 C_0) v^{26}
-(25509168 C_0+3520265184 \he_0^2) v^2-(2133532224 \he_0^2+2799360 C_0) v^{22}
\\
&-(127545840 C_0+21070572768 \he_0^2) v^6+23914845\,,
\end{split}
\eqlabel{resr5}
\end{equation}
\begin{equation}
\begin{split}
&\calr_6=-885735+(236196 C_0+62355744 \he_0^2) v^2-885735 v^4+(866052 C_0\\
&+288159120 \he_0^2) v^6
-2302911 v^8+(-26453952 \he_0^2-157464 C_0) v^{10}
-177147 v^{12}
\\
&+(66624768 \he_0^2-134136 C_0) v^{14}
+275562 v^{16}
+(-5412096 \he_0^2+27216 C_0) v^{18}
\\
&+92826 v^{20}
+(9072 C_0-1594080 \he_0^2) v^{22}
+7938 v^{24}
-(4968 C_0+1684800 \he_0^2) v^{26}\\
&-2214 v^{28}+(-31104 \he_0^2-648 C_0) v^{30}
-963 v^{32}+(-50976 \he_0^2+396 C_0) v^{34}
-531 v^{36}\\
&+(-1584 \he_0^2+12 C_0) v^{38}-3 v^{40}+v^{44}\,,
\end{split}
\eqlabel{resr6}
\end{equation}
\begin{equation}
\begin{split}
&\calr_7=7 v^{28}-3 v^{24}-93 v^{20}-63 v^{16}+1269 v^{12}-4617 v^8+6561 v^4+2187\,,
\end{split}
\eqlabel{resr7}
\end{equation}
\begin{equation}
\begin{split}
&\calr_8=1750802750915778 v^{16}
-547909426568250 v^{12}+551849751221706 v^8\\
&-3229924616793 v^4+10220357329866 v^{44}
+252995054673834 v^{32}\\
&+32427971352054 v^{40}+94323062131416 v^{36}+1924492298742 v^{48}
+491409014598 v^{52}\\
&+392340630598470 v^{28}+310374023297562 v^{20}+903540325216554 v^{24}
\\
&+31603772430 v^{56}+11854881954 v^{60}-177976890 v^{64}+55953441 v^{68}-376 v^{76}\\
&+185736 v^{72}
+16 v^{80}+(-77182514538916608 \he_0^2-361083701553408 C_0) v^{14}\\
&+(6169681384704 C_0+20228383300939776 \he_0^2) v^{26}
+(-39822513408 \he_0^2\\
&+685117440 C_0) v^{62}+(117792358596864 C_0+19949787269471232 \he_0^2) v^{10}\\
&+(211985675693568 C_0+64589051749911552 \he_0^2) v^{18}-(23950080 C_0+403273728 \he_0^2)\\
&\times  v^{66}
-(4040529154973568 \he_0^2+29968782786432 C_0) v^6+(8159616 \he_0^2+746496 C_0) v^{70}
\\&
+(24405013822464 \he_0^2-7899545088 C_0) v^{50}+(107114016798720 \he_0^2\\
&+297538935552 C_0) v^2+(-281691198720 C_0+348792330451968 \he_0^2) v^{42}\\
&+(20993178240 C_0-6791658865920 \he_0^2) v^{54}+(-3560308992 C_0+487741077504 \he_0^2) v^{58}\\
&+(-7095537792 C_0-2444692311194112 \he_0^2) v^{38}
+(2463432321024 C_0\\
&+3589000230500352 \he_0^2) v^{34}+(-15313028838912 C_0-16874613168627456 \he_0^2) v^{30}\\
&+(-177353046892800 \he_0^2+190336884480 C_0) v^{46}-(138461655146112 C_0\\
&+66703936055747328 \he_0^2) v^{22}+697356880200\,.
\end{split}
\eqlabel{resr8}
\end{equation}

For the computation of $C$ we actually need only the asymptotic solution of $\ha_3'(v)$ as $x\equiv 3^{1/4}-v\to 0_+$.
Using \eqref{haorder3}-\eqref{resr8} we find:
\begin{equation}
\begin{split}
&\frac{d\ha_3(v)}{dv}\bigg|_{x\equiv 3^{1/4}-v\to 0_+}=\frac{40\he_0 (-18 \he_0^2+\sqrt3)}{3^{3/4}}   \frac{1}{x^6}
-\frac{\he_0}{48} \biggl(256 C_0-512 \sqrt3 \ln2+1760 \sqrt3\\
&-(6432+3 \delta_1) \he_0^2\biggr) \frac{1}{x^5}
+\frac{3^{1/4} \he_0}{1152}  \biggl(-18432 \ln2-12096-12 \delta_1+\sqrt3 C_0 (3360+\delta_1)\\
&-24 \he_0 \sqrt3 \delta_2
+24 \sqrt3 (15792+5 \delta_1) \he_0^2\biggr) \frac{1}{x^4}
+\biggl(\frac{1}{12} 3^{1/4} (288+\delta_1) \cala_3+\frac{1}{5184} \sqrt3 (38880 \he_0\\
&-2 \delta_2+9 \he_0 \delta_1) C_0
+\frac{1}{96} \sqrt3 (67872+13 \delta_1) \he_0^3-\frac{17}{432} \sqrt3 \delta_2 \he_0^2
+\biggl(\frac{1025}{24}-46 \ln2\\
&+\frac{1}{36} \sqrt3 \delta_3+\frac{3475}{256}\pi\biggr) \he_0+\frac{1}{216} \delta_2\biggr) \frac{1}{x^3}+\hat{\cala}_3\ \frac{1}{x^2}
+\calo\left(x^{-1}\right)\,,
\end{split}
\eqlabel{expha3}
\end{equation}
where $\hat{\cala}_3$ must be fixed so to satisfy the boundary condition \eqref{bbcc}.

\section{Source terms for the framework of Kinoshita {\em et al}}

Order - 0:
\begin{eqnarray}
S_A^0&=& 180 y^{10} \left(16-23   y^4\right) \,,\nonumber \\
S_B^0&=&-4320 y^{10} \nonumber \,,\\
S_C^0&=&0 \,.
\end{eqnarray}
Order - 1:
\begin{eqnarray}
S_A^1&=&-\left(33 w \left(y^3+y^2+y+1\right)^2\right)^{-1} \nonumber \\
&& 4 y^6 \left(11 A_1^0 \left(y^3+2 y^2+3 y+4\right)-3 y^4
   \left(74945 y^{11}+191854 y^{10}+308763 y^9\right. \right.\nonumber \\
   && 
   +425672 y^8+370056
   y^7+209256 y^6+48456 y^5-112344 y^4-153504 y^3\nonumber \\
   && \left. \left. -109728 y^2-65952
   y-22176\right)\right)  \nonumber\,, \\
 S_B^1&=& \frac{8640 y^{10} (11+26 y)}{11 w} \nonumber \,,\\
S_C^1&=& -(33 w (y-1)
   \left(y^3+y^2+y+1\right)^3)^{-1} \nonumber \\
   &&  \left(2 y^3 \left(3 \left(84 y^{10}+168 y^9+252 y^8+2541 y^7-754
   y^6-4049 y^5-7344 y^4\nonumber \right. \right. \right. \\
   && \left. \left. \left. -15624 y^3-12528 y^2-9432 y-6336\right) y^7+11
   A_1^0 \left(3 y^2+2 y+1\right)\right)\right)\,.
\end{eqnarray}

\end{document}